\documentclass[12pt]{article}
\usepackage{cite}

\makeatletter
\def\eqnarray{ \stepcounter{equation} \let\@currentlabel=\theequation
 \global\@eqnswtrue
 \global\@eqcnt\z@
 \tabskip\@centering
 \let\\=\@eqncr
 $$\halign to \displaywidth\bgroup\@eqnsel\hskip\@centering
 $\displaystyle\tabskip\z@{##}$&\global\@eqcnt\@ne
 \hfil$\displaystyle{{}##{}}$\hfil
 &\global\@eqcnt\tw@$\displaystyle\tabskip\z@{##}$\hfil
 \tabskip\@centering&\llap{##}\tabskip\z@\cr}
\makeatother

\makeatletter
\def\@arrayacol{\edef\@preamble{\@preamble \hskip .2\arraycolsep}}
\def\array{\let\@acol\@arrayacol \let\@classz\@arrayclassz
\let\@classiv\@arrayclassiv \let\\\@arraycr\def\@halignto{}\@tabarray}
\makeatother
\renewcommand{\arraystretch}{1.6}
\begin{document}

\renewcommand{\theequation}{\arabic {section}.\arabic{equation}} %
\setlength{\baselineskip}{7mm} 
\begin{titlepage}
\begin{flushright}
EPHOU-00-011 \\
October, 2000
\end{flushright}
 
\vspace{15mm}

\begin{center} 
{\Large Generalized Gauge Theories and Weinberg-Salam Model} \\
{\Large  with Dirac-K\"ahler Fermions} \\
\vspace{1cm}
{\bf {\sc Noboru Kawamoto}}$^\dagger$, 
{\bf {\sc Takuya Tsukioka}}$^\ddagger$ 
{\bf {\sc and}} 
{\bf {\sc Hiroshi Umetsu}}$^\dagger$ \\
\vspace{3mm}

$\dagger${\it Department of Physics, Hokkaido University,}\\
{\it Sapporo, 060-0810, Japan}\\ 
kawamoto, umetsu@particle.sci.hokudai.ac.jp \\ 
\vspace{3mm}

$\ddagger${\it School of Theoretical Physics, 
           Dublin Institute for Advanced Studies,}\\
{\it 10 Burlington Road, Dublin, 4, Ireland}\\
tsukioka@stp.dias.ie 
\end{center}

\vspace{2cm}

\begin{abstract}
We extend previously proposed generalized gauge theory formulation of 
Chern-Simons type and topological Yang-Mills type actions into Yang-Mills 
type actions. 
We formulate gauge fields and Dirac-K\"ahler matter fermions by 
all degrees of differential forms. 
The simplest version of the model which includes only zero and one form gauge 
fields accommodated with the graded Lie algebra of $SU(2|1)$ 
supergroup leads Weinberg-Salam model. 
Thus the Weinberg-Salam model formulated by noncommutative geometry is 
a particular example of the present formulation. 
\end{abstract}

\end{titlepage}

%%%%%%%%%%%%%%%%%%%%
\section{Introduction}

\setcounter{equation}{0} 
\setcounter{footnote}{0}

It is obviously the most challenging question how we can formulate the
standard model together with quantum gravity in a unified way. 
The superstring related topics are motivated to challenge on this question. 
It is, however, not obvious that this is the unique way to find a clue to this
question.

Empirically to say lattice theories are successful to describe the quantum
and non-perturbative aspects of gauge field theories. The lattice QCD is one
example and two-dimensional quantum gravity is another successful example.
In the standard model there are several parameters; quark and lepton masses,
weak mixing angles, $\cdots$, which, we hope, would be numerically evaluated
by a quantitative description of the unified theory. We believe that the
lattice theory might again play an important role in the quantitative
formulation of the unified theory.

In formulating a gauge theory on a simplicial lattice manifold, we naively
expect that we should formulate the gauge theory by differential forms since
general coordinate invariance can be easily accommodated by differential
forms. Furthermore the form degree corresponds to the dimensions of simplex 
on the simplicial lattice manifold and thus $n$-form field variable may 
be assigned on the $n$-simplex of the simplicial lattice manifold.

One of the present authors (N. K.) and Watabiki proposed the generalized
gauge theory formulation for Chern-Simons type actions and topological
Yang-Mills actions which include all the degrees of differential forms yet
has the same algebraic structure as the ordinary gauge theory~\cite
{KW1,KW2,KW3}. The field variables are quaternion valued which classifies
bosonic even, bosonic odd, fermionic even, and fermionic odd forms. $Z_2$
grading structure for the field variables is a natural consequence of the
quaternion structure and thus the graded Lie algebra naturally comes in as a
gauge algebra. There are similar formulations related with the generalized
gauge theory~\cite{MP,CF}.

We can then expect that this type of the generalized gauge theory
formulation may provide a gauge theory formulation with gravity on the
simplicial lattice. In fact three-dimensional lattice gravity and
four-dimensional $BF$ lattice gravity have been formulated by using the
leading terms of the generalized Chern-Simons action which are the standard
Chern-Simons action in three dimensions and $BF$ action in four
dimensions~\cite{KNS,KSU}. These experiences of the formulation of 
lattice gravity including 0-, 1- and 2-form field variables provide 
us a feeling that the form variables may play an important role 
in the formulation of lattice gravity.

It has been clarified that the quantization of the generalized Chern-Simons
action is highly nontrivial, in fact infinitely reducible~\cite
{KOS,KSTU1,KSTU2}. The quantized minimal action has the same generalized
Chern-Simons type structure as the classical one. This would mean that we
have quantized the topological gravity in two dimensions and the topological
conformal gravity in four dimensions which were classically formulated by
the even-dimensional version of the generalized gauge theories~\cite{KW4,KW5}.

Connes pointed out that the Weinberg-Salam model can be formulated as a
particular case of the noncommutative geometry formulation 
of a gauge theory~\cite{Connes1,Connes2,Connes3}. 
Consider a manifold which is composed of a direct product of discrete two
points and four-dimensional flat space, $Z_2\times M_4$, and define a
connection and differential operator on this manifold. Due to the discrete
nature of the two points the differential operator can be represented by two
by two matrix. Thus the connection or equivalently the gauge field is now
represented by two by two matrix as well and thus possesses diagonal and
off-diagonal components. Then the weak and electromagnetic 1-form gauge
fields are assigned to the diagonal component while the 0-form Higgs
fields are assigned to the off-diagonal components. Then the spontaneously
broken Weinberg-Salam model comes out naturally from the pure Yang-Mills
action on this manifold by taking the group $SU(2)\times U(1)$~\cite
{Connes1,Connes2,Connes3}. This type of noncommutative geometry formulation
of Weinberg-Salam model have been intensively investigated. Here we give
partial lists of those investigations and the further references are therein
~\cite{Coq1,Coq2,Coq3,Coq4,CFF,NC1,NC2,NC3,NC4,NC5,NC6,NC7,NC8,NC9}.

We show that the two by two matrix representation of the gauge fields are
easily accommodated by the quaternions of the generalized gauge theory since
the quaternion algebra can be represented by two by two matrices. We can,
however, point out that our generalized gauge theory is more general
formulation as noncommutative geometry since it includes not only 0- and
1-form of gauge fields but also all the possible bosonic and fermionic
form degrees of gauge fields and gauge parameters and it can accommodate a
graded Lie algebra naturally. In fact we show in this paper that the graded
Lie algebra of $SU(2|1)$ supergroup leads naturally to the Weinberg-Salam
model. The importance of the $SU(2|1)$ graded Lie algebra in connection with
the standard model was first pointed out by Ne'eman~\cite{Neeman}. Later the
noncommutative geometry formulation of $SU(2|1)$ graded Lie algebra were
given by Coquereaux et al.~\cite{Coq2,Coq3,Coq4}. We believe that our
formulation of the Weinberg-Salam model will provide new insights into the
formulation of the standard model.

Witten pointed out that four-dimensional $N=2$ super Yang-Mills action 
comes out from the topological Yang-Mills action with instanton gauge 
fixing via twisting mechanism~\cite{Witten}. 
The quantization of the generalized topological Yang-Mills action 
in two dimensions with instanton gauge fixing was investigated. 
It was clarified that $N=2$ super Yang-Mills action comes out
naturally and the matter fermions appears from ghosts of quantization via
twisting mechanism~\cite{KT}. The twisting mechanism and the
Dirac-K\"ahler fermion formulation~\cite{DK1,DK2,DK3} are essentially
related through $N=2$ supersymmetry.

In the lattice gauge theory there is the well known chiral fermion problem.
The staggered fermion formulation~\cite{KS} made it clear that the
Kogut-Susskind fermion formulation~\cite{Susskind1,Susskind2} can avoid the
problem in such a way that there appear 4 copies of flavor suffices. The
curved space version of the Susskind fermion formulation or equivalently
the staggered fermion formulation on the flat spacetime is the
Dirac-K\"ahler fermion formulation which is formulated by differential
forms~\cite{DK1,DK2,DK3,BJ1,BJ2,BJ3,BJ4,BJ5,BJ6}. 
In formulating matter fermions
in the Weinberg-Salam model, we employ the Dirac-K\"ahler fermion
formulation so that all the gauge fields and matter fermions are formulated
by the differential forms. This would provide us a hope that the standard
model with matter fermion coupled to the gravity could be formulated purely
by differential forms and thus leads to a unified model with gravity on the
simplicial lattice. This kind of non-standard overview on unifying the
standard model with gravity on the lattice was reviewed in~\cite{Kawamoto}.

This paper is organized as follows: We summarize the generalized gauge
theory formulation of Chern-Simons type and topological Yang-Mills type
actions in section 2. We then extend the generalized gauge theory
formulation into Yang-Mills type action in section 3. In section 4 we
provide a generalized theory version of the Dirac-K\"ahler fermion
formulation. We then formulate the Weinberg-Salam model by the generalized
Yang-Mills action and Dirac-K\"ahler fermion formulation with $SU(2|1)$
graded Lie algebra in section 5. Summary and discussions will be given in
the last section.

%%%%%%%%%%%%%%%%%%%%
\section{Generalized gauge theory in arbitrary dimensions}

\setcounter{equation}{0} 
\setcounter{footnote}{0}

The generalized Chern-Simons actions, which were proposed by one of the
present authors and Watabiki about ten years ago, is a generalization of the
ordinary three-dimensional Chern-Simons theory into arbitrary dimensions~
\cite{KW1,KW2}. The essential point of the generalization is to extend a
1-form gauge field and 0-form gauge parameter to a quaternion valued
generalized gauge field and gauge parameter which contain all the possible
degrees of differential forms. Correspondingly the standard gauge symmetry
is extended to much higher topological symmetry. These generalizations are
formulated in such a way that the generalized actions have the same
algebraic structure as the ordinary three-dimensional Chern-Simons action.

Since this generalized Chern-Simons action can be formulated completely
parallel to the ordinary gauge theory, the generalization can be extended
further to the topological Yang-Mills actions. The generalized Chern-Simons
actions and topological Yang-Mills actions have topological nature from the
construction of the actions themselves. This could naively be understood
from the fact that in the generalized gauge theory formulation the
generalized gauge fields and gauge parameters contain exactly the same
number of field variables and thus all the generalized gauge fields could be
gauged away. If we, however, try to construct generalized Yang-Mills
actions, the construction does not proceed completely parallel to the
generalized Chern-Simons actions and topological Yang-Mills actions since
the topological nature is lost in the construction of the generalized
Yang-Mills action. It is one of the main subjects of this paper to formulate
the generalized Yang-Mills actions together with matter fermions in terms of
differential forms. Before getting into the details of formulating the
generalized Yang-Mills action we summarize the results of generalized gauge
theory for Chern-Simons type and topological Yang-Mills type actions.

In the most general form, a generalized gauge field $\mathcal{A}$ and a
gauge parameter $\mathcal{V}$ are defined by the following component form: 
\begin{eqnarray}
\mathcal{A} & = & {\mbox{\bf 1}}\psi + {\mbox{\bf i}} \hat{\psi} + 
{\mbox{\bf j}} A + {\mbox{\bf k}} \hat{A},  
\label{eqn:ggf} \\
\mathcal{V} & = & {\mbox{\bf 1}} \hat{a} + {\mbox{\bf i}} a + {\mbox{\bf j}} 
\hat{\alpha} + {\mbox{\bf k}} \alpha,  
\label{ggp0}
\end{eqnarray}
where $( \psi, \alpha )$, $( \hat{\psi}, \hat{\alpha} )$, $( A, a )$ and $( 
\hat{A},\hat{a} )$ are direct sums of fermionic odd forms, fermionic even
forms, bosonic odd forms and bosonic even forms, respectively, and they take
values on a gauge algebra. The bold face symbols $\mathbf{1}$, $\mathbf{i}$, 
$\mathbf{j}$ and $\mathbf{k}$ satisfy the algebra 
\begin{equation}
\begin{array}{c}
\mathbf{1}^2=\mathbf{1}, \quad \mathbf{i}^2=\epsilon_1 \mathbf{1}, \quad 
\mathbf{j}^2=\epsilon_2 \mathbf{1}, \quad \mathbf{k}^2=-\epsilon_1
\epsilon_2 \mathbf{1}, \\ 
\mathbf{i}\mathbf{j}=-\mathbf{j}\mathbf{i}=\mathbf{k}, \quad \mathbf{j}
\mathbf{k}=-\mathbf{k}\mathbf{j}=-\epsilon_2 \mathbf{i}, \quad \mathbf{k}
\mathbf{i}=-\mathbf{i}\mathbf{k}=-\epsilon_1 \mathbf{j},
\end{array}
\label{quaternion}
\end{equation}
where $(\epsilon_1,\epsilon_2)$ takes the value $(-1,-1), (-1,+1), (+1,-1)$
or $(+1,+1)$. We may call this algebra as ``quaternion algebra''.

The components of the gauge field $\mathcal{A}$ and parameter $\mathcal{V}$
are assigned to the elements of the gauge algebra in a specific way: 
\begin{equation}
\begin{array}{rclrclrclrcl}
A & = & T_a A^a, \quad \hat{\psi} & = & T_a \hat{\psi}^a, \quad \psi & = & 
\Sigma_\alpha \psi^\alpha, \quad \hat{A} & = & \Sigma_\alpha \hat{A}^\alpha,
&  &  &  \\ 
\hat{a} & = & T_a \hat{a}^a, \quad \alpha & = & T_a \alpha^a, \quad 
\hat{\alpha} & = & \Sigma_\alpha \hat{\alpha}^\alpha, \quad a 
& = & \Sigma_\alpha a^\alpha. &  &  & 
\end{array}
\end{equation}
The following graded Lie algebra can be adopted as a gauge algebra:
\begin{equation}
\left[ T_a , T_b \right] = f^c_{ab} T_c, \quad 
\left[ T_a , \Sigma_\beta \right] = g^\gamma_{a\beta} \Sigma_\gamma, \quad
\left\{ \Sigma_\alpha , \Sigma_\beta \right\} = h^c_{\alpha\beta} T_c,  
\label{gla1}
\end{equation}
where $[ \ , \ ]$ and $\{ \ , \ \}$ are commutator and anticommutator
respectively. All the structure constants are subject to consistency
conditions which follow from the graded Jacobi identities. If we choose 
$\Sigma_\alpha=T_a$ especially, this algebra reduces to 
$T_a T_b = k^c_{ab}T_c$ which is closed under multiplication. 
A specific example of such algebra is realized by Clifford algebra~\cite{KW4}.

An element having the same type of component expansion as $\mathcal{A}$ or 
$\mathcal{V}$ belongs to $\Lambda_-$ or $\Lambda_+$ class, respectively, 
\begin{eqnarray}
\lambda_- & = & {\mbox{\bf 1}}{\mbox{(fermionic odd)}} 
+ {\mbox{\bf i}} {\mbox{(fermionic even)}}  \nonumber \\
&+& {\mbox{\bf j}} {\mbox{(bosonic odd)}} + {\mbox{\bf k}} {\mbox{(bosonic
even)}} \quad \in \Lambda_-,  
\label{lambdaminus} \\
\lambda_+ & = & {\mbox{\bf 1}}{\mbox{(bosonic even)}} 
+ {\mbox{\bf i}}{\mbox{(bosonic odd)}}  \nonumber \\
&+& {\mbox{\bf j}}{\mbox{(fermionic even)}} + 
{\mbox{\bf k}}{\mbox{(fermionic odd)}} \quad \in \Lambda+.  
\label{lambdapulus}
\end{eqnarray}
These elements fulfill the following $Z_2$ grading structure: 
$$
[\lambda_+ , \lambda_+] \in \Lambda_+, \quad [\lambda_+ , \lambda_-] \in
\Lambda_-, \quad \{\lambda_- , \lambda_-\} \in \Lambda_+.
$$
The elements of $\Lambda_-$ and $\Lambda_+$ can be regarded as
generalizations of odd forms and even forms, respectively. 
The generalized exterior derivative which belongs to $\Lambda_-$ is given by 
\begin{equation}
Q=\mathbf{j}d \quad \in \Lambda_-,  
\label{gdo}
\end{equation}
and the following graded Leibniz rule similar to the ordinary differential
algebra holds: 
\begin{equation}
\{\vec{Q} , \lambda_-\}=Q\lambda_-, \quad [\vec{Q} , \lambda_+]=Q\lambda_+,
\quad Q^2=0,  
\label{graded leibniz rule 1}
\end{equation}
where $\lambda_+ \in \Lambda_+$ and $\lambda_- \in \Lambda_-$. The arrow of
the differential operator denotes $\vec{Q}\lambda_-=Q\lambda_--\lambda_-Q$.
In the ordinary differential algebra of exterior derivative with $Q$
replaced by the exterior derivative $d$, $\Lambda_-$ and $\Lambda_+$ are odd
and even form variables, respectively.

To construct the generalized Chern-Simons and topological Yang-Mills actions, 
we need to introduce two kinds of traces satisfying 
\begin{equation}
\begin{array}{rclcrcl}
\mbox{Tr}[T_a , \cdots] &=& 0, & {\qquad} 
& \mbox{Tr}[\Sigma_\alpha, \cdots] &=& 0, \\ 
\mbox{Str}[T_a , \cdots] &=& 0, & {\qquad} 
& \mbox{Str}\{\Sigma_\alpha, \cdots\} &=& 0,
\end{array}
\label{eqn:gtr}
\end{equation}
where $(\cdots)$ in the commutators or the anticommutators denotes a product
of generators. In particular $(\cdots)$ should include an odd number of 
$\Sigma_\alpha$'s in the last equation of (\ref{eqn:gtr}). $\mbox{Tr}$ is the
usual trace while $\mbox{Str}$ is the super trace satisfying the above
relations. We can then derive the following relations: 
\begin{eqnarray}
\mbox{Tr}_{\mbox{\bf k}}[\lambda_+,\lambda_-]&=& 
\mbox{Tr}_{\mbox{\bf k}}[\lambda_+,{\lambda^{\prime}}_+] =0,  \nonumber \\
\mbox{Str}_{\mbox{\bf j}}[\lambda_+,\lambda_-]&=& \mbox{Str}_{\mbox{\bf j}}
[\lambda_+,{\lambda^{\prime}}_+] =0,  \\
\mbox{Str}_{\mbox{\bf 1}}[\lambda_+,{\lambda^{\prime}}_+]&=& 
\mbox{Str}_{\mbox{\bf 1}}\{ \lambda_-,{\lambda^{\prime}}_-\} =0, \nonumber
\label{graded component relation}
\end{eqnarray}
where $\mbox{Tr}_{\mathbf{q}}(\cdots)$ and $\mbox{Str}_{\mathbf{q}}(\cdots)
\ (\mathbf{q}=\mathbf{1}, \mathbf{j}, \mathbf{k})$ are defined so as to
pick up only the coefficients of $\mathbf{q}$ from $(\cdots)$ and take the
traces defined by eq.(\ref{eqn:gtr}). These definitions of the traces will
be crucial to show that the generalized gauge theory action can be invariant
under the generalized gauge transformation bellow.

As we have seen in the above the generalized gauge field $\mathcal{A}$ and
parameter $\mathcal{V}$, and the generalized differential operator $Q$ play
the same role as the 1-form gauge field, 0-form gauge parameter and
differential operator of the usual gauge theory, respectively. We can then
construct generalized actions in terms of these generalized quantities. We
first define a generalized curvature 
\begin{equation}
\mathcal{F} \equiv \frac{1}{2}\{Q+\mathcal{A},Q+\mathcal{A}\} = Q\mathcal{A}
+ \mathcal{A}^2.  
\label{eqn:gcurv}
\end{equation}

We can construct the generalized actions of Chern-Simons type which have the
standard form with respect to the generalized quantities 
%
%%%%%
\renewcommand{\arraystretch}{2.0}
%%%%%
%
\begin{equation}
\begin{array}{rcl}
\displaystyle{S^e_{GCS}} & = & 
\displaystyle{\int_M \mbox{Tr}_{\mbox{\bf k}} 
\left( \frac{1}{2}\mathcal{A}Q\mathcal{A}+\frac{1}{3}\mathcal{A}^3 \right)}, \\
\displaystyle{S^o_{GCS}} & = & 
\displaystyle{\int_M \mbox{Str}_{\mbox{\bf j}} 
\left( \frac{1}{2}\mathcal{A}Q\mathcal{A}+\frac{1}{3}\mathcal{A}^3 \right)}, 
\end{array}
\label{gcsa}
\end{equation}
%
%%%%%
\renewcommand{\arraystretch}{1.6}
%%%%%
%
where $\displaystyle{S^e_{GCS}}$ is even-dimensional generalized
Chern-Simons action while $\displaystyle{S^o_{GCS}}$ is odd-dimensional
generalized Chern-Simons action. When we consider $D$-dimensional manifold 
$M$, we need to pick up $D$-form terms in the action.

Similarly we can obtain the generalized topological Yang Mills action by
taking the $\mathbf{1}$-th component which is even-dimensional counterpart
of the bosonic component, 
\begin{equation}
\displaystyle{S_{GTYM}=\int_M \mbox{Str}_{\mbox{\bf 1}} \left( \mathcal{F} 
\mathcal{F} \right)}.  
\label{gtyma}
\end{equation}

In these generalized actions we have given the examples of taking the 
$\mathbf{k}$, $\mathbf{j}$, $\mathbf{1}$-th component to pick up 
the even- and odd-dimensional 
counter part of bosonic generalized Chern-Simons actions and even-dimensional
counterpart of bosonic generalized topological Yang-Mills action,
respectively. We can, however, construct all the different types; bosonic
even, bosonic odd, fermionic even, fermionic odd, Chern-Simons type and
topological Yang-Mills type actions by taking $\mathbf{q}=\mathbf{1}$,  
$\mathbf{i}$, $\mathbf{j}$, $\mathbf{k}$-th components of 
$\mbox{Tr}_{\mathbf{q}}(\cdots)$ and $\mbox{Str}_{\mathbf{q}}(\cdots)$.

Those generalized Chern-Simons actions (\ref{gcsa}) and the topological
Yang-Mills actions (\ref{gtyma}) are invariant under the following
generalized gauge transformations: 
\begin{equation}
\delta \mathcal{A}=[Q+\mathcal{A} , \mathcal{V} ],  
\label{ggt1}
\end{equation}
where $\mathcal{V}$ is the generalized gauge parameter defined by eq.(\ref
{ggp0}). Correspondingly the generalized gauge transformation of the
generalized curvature (\ref{eqn:gcurv}) is given by 
\begin{equation}
\delta \mathcal{F}=[\mathcal{F} , \mathcal{V} ].  
\label{eqn:gtgcurv}
\end{equation}
It should be noted that this symmetry is much larger than the usual gauge
symmetry, in fact topological symmetry, since the gauge parameter 
$\mathcal{V}$ contains as many gauge parameters as gauge fields of various 
forms in $\mathcal{A}$.

We now show the explicit form of these generalized actions. Firstly we
define the explicit form of the generalized gauge fields and gauge
parameters by 
\begin{eqnarray*}
{\mathcal{A}} &=& \mbox{\bf{j}}A + \mbox{\bf{k}}\hat{A}  \\
&\equiv& \mbox{\bf{j}}(\omega^{a} + \Omega^{a} + \cdots)T_{a} + \mbox{\bf{k}}
(\phi^{\alpha} + B^{\alpha} + H^{\alpha} + \cdots)\Sigma_{\alpha},
\label{ggf1} \\
{\mathcal{V}} &=& \mbox{\bf{1}}\hat{a} + \mbox{\bf{i}}a  \\
&\equiv& \mbox{\bf{1}}(v^{a} + b^{a} + V^{a}+ \cdots)T_{a} + \mbox{\bf{i}}
(u^{\alpha} + U^{\alpha} + \cdots)\Sigma_{\alpha},
\label{ggp1}
\end{eqnarray*}
where $\phi$, $\omega$, $B$, $\Omega$, $H$ are bosonic 0-, 1-, 2-, 3-,
4-form gauge fields and $v,u,b,U,V$ are bosonic 0-, 1-, 2-, 3-, 4-form gauge
parameters, respectively. We have omitted the fermionic degrees for
simplicity. The generators $T_{a}$ and $\Sigma_{\alpha}$ satisfy the graded
Lie algebra (\ref{gla1}).

We show concrete expressions of the generalized Chern-Simons actions in two,
three and four dimensions where we choose particular quaternion algebra 
$(\epsilon_1,\epsilon_2)=(-1,-1)$ in this section for simplicity, 
\begin{eqnarray*}
S_2 &=& - \int \mbox{Tr}\{\phi(d\omega + \omega^2) + \phi^2 B\}, \\
S_3 &=& - \int \mbox{Str} \left\{ \frac{1}{2}\omega d\omega + \frac{1}{3}
\omega^3 - \phi(d B + [\omega,B]) + \phi^2 \Omega \right\}, \\
S_4 &=& - \int \mbox{Tr}\{ B(d\omega + \omega^2) + \phi(d \Omega + \{\omega,
\Omega\}) + \phi B^2 + \phi^2 H \},
\end{eqnarray*}
which are invariant under the following gauge transformations: 
\begin{eqnarray*}
\delta \phi &=& [\phi,v],  \\
\delta \omega &=& dv + [\omega,v] - \{\phi,u\},   \\
\delta B &=& du + \{\omega,u\} + [B,v] + [\phi,b],  \\
\delta \Omega &=& db + [\omega,b] + [\Omega,v] - \{B,u\} + \{\phi,U\}, \\ 
\delta H &=& -dU-\{\omega,U\}+\{\Omega,u\}+[H,v]+[B,b]+[\phi,V].  
\end{eqnarray*}

In ordinary gauge theory the integral for the trace of the $n$-th power of
curvature is called $n$-th Chern character and has topological nature. In the
generalized gauge theory it is possible to define generalized Chern
character which is expected to have topological nature related to
``generalized index theorem"
\begin{eqnarray}
{\mbox{Str}}_{\mbox{\bf 1}}({\mathcal{F}}^n) ={\mbox{Str}}_{\mbox{\bf 1}}
(Q\Omega_{2n-1}),   
\label{eqn:bopo}
\end{eqnarray}
where $\Omega_{2n-1}$ is the ``generalized" Chern-Simons forms.
Especially, for $n=2$ case in (\ref{eqn:bopo}), we obtain the topological
Yang-Mills type action of (\ref{gtyma}) on an even-dimensional manifold $M$
related to the generalized Chern-Simons action with one dimension lower, 
\begin{equation}
\int_{M}{\mbox{Str}}_{\mbox{\bf 1}}\Big(\frac{1}{2}\mathcal{F}^2\Big) 
=\int_{M}{\mbox{Str}}_{\mbox{\bf 1}}\Bigg( Q\Big(\frac{1}{2}{\mathcal{A}}Q{
\mathcal{A}} +\frac{1}{3}\mathcal{A}^3\Big)\Bigg),  
\label{eqn:2cc}
\end{equation}
which has the same form of the standard relation. The generalized gauge
theory version of this topological relation (\ref{eqn:2cc}) in four
dimensions can be explicitly given by 
\begin{eqnarray*}
&&\int_{M_4}{\mbox{Str}}\Big[ \frac{1}{2}(d\omega + \omega^2)^2 +(d\omega+
\omega^2)\{\phi,B \}+ \phi^2(d\Omega + \{\omega,\Omega\})  \nonumber \\
&&\hspace{14mm} -(d\phi + [\omega, \phi])(dB + [\omega, B]+[\Omega, \phi]) 
   + \frac{1}{2}B^4 + \frac{1}{2}\{\phi, B \}^2 +\phi^2B^2)\Big]  \nonumber \\
&=&\int_{M_4}{\mbox{Str}}d\Big[\frac{1}{2}\omega d \omega + \frac{1}{3}
\omega^3-\phi(dB+[\omega,B])+\phi^2\Omega \Big],
\end{eqnarray*}
where the first term in the left and right hand side includes the ordinary
2-form curvature square term and Chern-Simons term. This relation, however,
includes all the degrees of differential forms up to 4-form, and yet
fulfills highly non-trivial topological relation which is expected to have
mathematical background.

%%%%%%%%%%%%%%%%%%%%

\section{Generalized Yang-Mills actions}

\setcounter{equation}{0}
\setcounter{footnote}{0}

In this section we formulate Yang-Mills action in terms of differential
forms. In particular we introduce higher degrees of differential forms as
gauge fields in contrast with the standard Yang-Mills action where only 
1-form gauge field is introduced. We have defined generalized gauge field in 
(\ref{eqn:ggf}) where all degrees of differential forms with bosonic and
fermionic antisymmetric tensor fields are introduced. For simplicity we omit
to introduce fermionic gauge fields in this section. We here again show the
explicit form of the generalized gauge field 
\begin{eqnarray}
{\mathcal{A}} &=& \mbox{\bf{j}}A + \mbox{\bf{k}}\hat{A}  \nonumber \\
&\equiv& \mbox{\bf{j}}(\omega^{a} + \Omega^{a} + \cdots)T_{a} + \mbox{\bf{k}}
(\phi^{\alpha} + B^{\alpha} + H^{\alpha} + \cdots)\Sigma_{\alpha}.
\label{ggf2}
\end{eqnarray}
It is important to realize here that the graded Lie algebra naturally comes
into the generalized gauge theory formulation from the beginning as we saw
in the previous section.

Here we generalize the differential operator (\ref{gdo}) in the following
form: 
\begin{eqnarray}
\mathcal{D} &=& \mbox{\bf{j}}d + \mbox{\bf{k}}m  \nonumber \\
&=& \mbox{\bf{j}}d + \mbox{\bf{k}}m^{\alpha}\Sigma_{\alpha},  
\label{gdo2}
\end{eqnarray}
where $d$ is the standard exterior derivative and $m$ is a constant matrix.
It should be noted that this newly defined generalized differential operator 
$\mathcal{D}$ stays to be an element of $\Lambda_-$ class and satisfies the
graded Leibniz rule similar to (\ref{graded leibniz rule 1}) 
\begin{equation}
\{\vec{\mathcal{D}} , \lambda_-\}=\mathcal{D}\lambda_-, \qquad 
[\vec{\mathcal{D}} , \lambda_+]=\mathcal{D}\lambda_+, \qquad  
\label{graded leibniz rule 2}
\end{equation}
where $\lambda_+ \in \Lambda_+$ and $\lambda_- \in \Lambda_-$. It should,
however, be noted that the nilpotency of $\mathcal{D}$ is not satisfied in
general since $\mathcal{D}^2=\mbox{\bf{k}}^2m^2$.

The products of the form-valued elements $\lambda_-$, $\lambda_+$ are 
meant to be the wedge product and we do not write the wedge $\wedge$ 
explicitly in the case of wedge product unless it is necessary 
to stress the difference from the cup product $\vee$ which will be 
defined shortly.

Here we define the following generalized curvature which is the naive
generalization of the generalized curvature (\ref{eqn:gcurv}): 
\begin{eqnarray}
\mathcal{F} &\equiv& \frac{1}{2} \{ \mathcal{D} + \mathcal{A}, \mathcal{D} + 
\mathcal{A} \} = \mathcal{D}^2 + \{ \mathcal{D}, \mathcal{A} \} 
+ \mathcal{A}^2  \nonumber \\
&\equiv& \epsilon_2 \Big[ \mathbf{1} (\mathcal{F}^{(0)} + 
\mathcal{F}^{(2)} + \mathcal{F}^{(4)} + \cdots ) + 
\mathbf{i} (\mathcal{F}^{(1)} + 
\mathcal{F}^{(3)} + \cdots ) \Big],   
\label{gc}
\end{eqnarray}
where 
%
%%%%%
\renewcommand{\arraystretch}{2.0}
%%%%%
%
\begin{equation}
\begin{array}{rcl}
\mathcal{F}^{(0)} &=& -\epsilon_1(\phi + m)^2, \\
\mathcal{F}^{(1)} &\equiv& \mathcal{F}^{(1)}_\mu dx^\mu = -d\phi
-[\omega,\phi+m],   \\
\mathcal{F}^{(2)} &\equiv& 
\displaystyle{\frac{1}{2!}}\mathcal{F}^{(2)}_{\mu\nu}
dx^\mu \wedge dx^\nu = d\omega + \omega^2 - \epsilon_1\{\phi + m, B \},\\
\mathcal{F}^{(3)} &\equiv& 
\displaystyle{\frac{1}{3!}}\mathcal{F}^{(3)}_{\mu\nu\rho}
dx^\mu \wedge dx^\nu \wedge dx^\rho = -dB - [\omega,B] - [\Omega,\phi + m], \\
\mathcal{F}^{(4)} &\equiv& 
\displaystyle{\frac{1}{4!}}\mathcal{F}^{(4)}_{\mu\nu\rho\sigma} 
dx^\mu \wedge dx^\nu \wedge dx^\rho \wedge dx^\sigma 
= d\Omega + \{\omega,\Omega \} - \epsilon_1 (B^2+\{\phi+m,H \}), \\
&\cdots& . 
\end{array}
\label{defofcurv}
\end{equation}
% 
%%%%%
\renewcommand{\arraystretch}{1.6}
%%%%%
%
Here in this definition it can be recognized that the addition of the
constant matrix to the differential operator simply shift the value of
constant term in the 0-form gauge field $\phi$.

In order to define generalized Yang-Mills action we introduce the notion of
Clifford product or simply cup product $\vee$ which was introduced by
K\"ahler~\cite{DK2}. 
This cup product $\vee$ should be differentiated from the wedge product 
$\wedge$. We consider an element $u$ which is the direct sum of forms 
\begin{eqnarray}
u= \sum^m_{p=0} \frac{1}{p!} u_{\mu_1\cdots\mu_p}
dx^{\mu_1}\wedge\cdots\wedge dx^{\mu_p},  
\label{gf}
\end{eqnarray}
where $m\le D$ with $D$ as spacetime dimension.

We then define the linear operator $e_{\mu}$, 
\begin{eqnarray}
e_\mu u= \sum^{m-1}_{p=0} \frac{1}{p!} u_{\mu\mu_1\cdots\mu_p}
dx^{\mu_1}\wedge\cdots\wedge dx^{\mu_p},  
\label{cop}
\end{eqnarray}
which is understood as a differentiation of the polynomial of differential
form with respect to $dx^\mu$. In particular it plays the role of
contracting operator as 
\begin{eqnarray*}
&&e_\mu dx^{\alpha_1}\wedge dx^{\alpha_2}\wedge dx^{\alpha_3}\wedge
dx^{\alpha_4} \wedge\cdots  \\
&=& g_\mu^{\alpha_1}dx^{\alpha_2}\wedge dx^{\alpha_3}\wedge dx^{\alpha_4}
\wedge\cdots - g_\mu^{\alpha_2}dx^{\alpha_1}\wedge dx^{\alpha_3} \wedge
dx^{\alpha_4} \wedge\cdots  \\
&+& g_\mu^{\alpha_3}dx^{\alpha_1}\wedge dx^{\alpha_2} \wedge dx^{\alpha_4}
\wedge\cdots - g_\mu^{\alpha_4}dx^{\alpha_1}\wedge dx^{\alpha_2} \wedge
dx^{\alpha_3} \wedge\cdots \\
&+& \cdots.  
%\label{copex}
\end{eqnarray*}

We can now define the cup product of $u$ and $v$, which have the
same expansion form as (\ref{gf}), by 
\begin{eqnarray}
u\vee v= \sum^m_{p=0} \frac{1}{p!} \zeta_p \{ \eta^p(e_{\mu_1}\cdots
e_{\mu_p} u) \} \wedge e^{\mu_1}\cdots e^{\mu_p} v,  
\label{cp}
\end{eqnarray}
where we need to introduce the sign operators $\eta$ and $\zeta$ which
generate the following sign factors: 
\begin{equation}
\begin{array}{rcl}
\zeta u_p &=& \zeta_p u_p \equiv (-1)^{\frac{p(p-1)}{2}} u_p,\\
\eta u_p &=& (-1)^p u_p,  
\end{array}
\label{sf}
\end{equation}
where $u_p$ is a $p$-form variable. These sign operators satisfy the following
properties: 
\begin{equation}
\begin{array}{rclcrcl}
\eta(u_p\wedge w_q) &=& (\eta u_p)\wedge (\eta w_q), &\quad& 
\eta(u_p\vee w_q) &=& (\eta u_p)\vee (\eta w_q),   \\
\zeta(u_p\wedge w_q) &=& (\zeta w_q)\wedge (\zeta u_p), &\quad& 
\zeta(u_p\vee w_q) &=& (\zeta w_q)\vee (\zeta u_p),  
\label{sfrelations}
\end{array}
\end{equation}
where $u_p$ and $w_q$ are arbitrary $p$- and $q$-form variables, respectively.
The sign factors $\zeta_p$ and the operator $\eta^p$ are necessary to ensure
the associativity of the cup product 
\begin{eqnarray}
(u\vee v)\vee w = u\vee (v\vee w). 
\label{associativity}
\end{eqnarray}

We now define the generalized Yang-Mills action in $D$ dimensions 
with $e$ as the coupling constant 
\begin{eqnarray}
S_G &=& \frac{1}{e^2}\int 
\hbox{Tr}\Big[{\mathcal{F}}\vee {\mathcal{F}}\Big]_{\hbox{{\bf 1}} } * 1  
\nonumber \\
&=& \frac{1}{e^2}\int d^Dx\sqrt{g} 
\hbox{Tr}\Big[\sum^D_{p=0} \zeta_p \frac{1}{p!}  
{\mathcal{F}}^{(p)}_{\mu_1\cdots \mu_p}{\mathcal{F}}^{(p)\mu_1\cdots
\mu_p}\Big],  
\label{gga}
\end{eqnarray}
where we have taken the particular choice $\epsilon_1=1$ which will be shown
as a natural choice later. The Hodge star $*$ acting on $1$ denotes the
invariant volume element. For example in four dimensions $* 1 = \sqrt{g}
\frac{1}{4!}\epsilon_{\alpha_1\alpha_2\alpha_3\alpha_4} dx^{\alpha_1}\wedge
dx^{\alpha_2}\wedge dx^{\alpha_3}\wedge dx^{\alpha_4} =\sqrt{g}d^4x$ with $g$
as the metric determinant. We need to pick up 0-form components in the trace
due to the presence of $* 1$ . The symbol $\mathbf{1}$ denotes to pick up
the coefficient of $\mathbf{1}$ for the quaternion expansion in the trace.
The reason of this particular choice of the component is due to the fact
that the coefficient of $\mathbf{1}$ is even form and thus includes 0-form
terms.

The generalized gauge parameter $\mathcal{V}$ includes all degrees of
differential forms as in (\ref{ggp0}) where we do not consider the fermionic
gauge parameters here. The generalized gauge field $\mathcal{A}$ with the
new definition of the generalized differential operator (\ref{gdo2})
transforms similarly as (\ref{ggt1}) 
\begin{equation}
\delta \mathcal{A}=[\mathcal{D}+\mathcal{A} , \mathcal{V} ].  
\label{ggt2}
\end{equation}
The gauge transformation of newly defined generalized curvature (\ref{gc})
leads 
\begin{equation}
\delta \mathcal{F}=\{\mathcal{D}+\mathcal{A} , \delta\mathcal{A} \} = 
[\mathcal{F},\mathcal{V}].  
\label{gtgc0}
\end{equation}

If we now consider the gauge invariance of the generalized Yang-Mills action
(\ref{gga}) under the generalized gauge transformation (\ref{ggt2}), we
notice that the gauge invariance is lost in general due to the loss of the
associativity since the wedge product and the cup product are mixed up in
the generalized gauge transformation of the action. There is, however, an
exception that the associativity is recovered when the gauge parameter
includes only a 0-form. This is the standard situation where the gauge
parameter normally includes 0-form only for the gauge transformation of the
Yang-Mills action. We thus define the generalized gauge parameter 
\begin{equation}
{\ \mathcal{V}}_0= \hbox{{\bf 1}} v^aT_a,  
\label{ggp zero}
\end{equation}
which includes only 0-form gauge parameter $v$. 
This should be compared with the generalized gauge parameter of generalized
Chern-Simons formulation where all the degrees of differential forms were
introduced. The generalized gauge transformation now leads to 
\begin{equation}
\delta {\mathcal{A}}= [{\mathcal{D}}+{\mathcal{A}},\mathcal{V}_0] = 
[{\mathcal{D}}+{\mathcal{A}},\mathcal{V}_0]_{\vee},  
\label{ggt3}
\end{equation}
where the first commutator is the standard commutator with wedge product
while the second commutator is defined as: 
$$
[\mathcal{A},\mathcal{B}]_\vee = \mathcal{A}\vee\mathcal{B}-
\mathcal{B}\vee\mathcal{A}.  
\label{cup product of commutator}
$$
The last equality of (\ref{ggt3}) is satisfied since ${\mathcal{V}}_0$
includes only 0-form. Then the gauge transformation of the generalized
curvature is given by 
\begin{eqnarray}
\delta {\mathcal{F}}= [{\mathcal{F}},{\mathcal{V}}_0] = [{\mathcal{F}}, 
{\mathcal{V}}_0]_{\vee},  
\label{gtgc1}
\end{eqnarray}
or equivalently 
\begin{eqnarray}
\delta {\mathcal{F}}^{(p)}= [{\mathcal{F}}^{(p)},v]_\vee.  
\label{gtgc2}
\end{eqnarray}
We can then show that the generalized Yang-Mills action $S_G$ is gauge
invariant under the generalized gauge transformation (\ref{ggt3}) since the
gauge transformation of the action can be written in the following form: 
\begin{eqnarray*}
\delta S_G &=& \frac{1}{e^2}\int \hbox{Tr}\Big[\{\delta{\mathcal{F}}, 
{\mathcal{F}}\}_\vee\Big]_{\hbox{{\bf 1}}}*1  \\
&=& \frac{1}{e^2}\int \sum^D_{p=0}\hbox{Tr}\Big[\{\delta {\mathcal{F}}^{(p)}, 
{\mathcal{F}}^{(p)}\}_\vee\Big]*1   \\
&=& 0.  
\label{gtga}
\end{eqnarray*}
In the proof of the gauge invariance the associativity of the cup product 
(\ref{associativity}) and the following relation should be used: 
$$
\hbox{Tr}[\lambda_+\vee\lambda^{\prime}_+]_{\hbox{{\bf 1}}} \ast 1 = 
\hbox{Tr}[\lambda^{\prime}_+\vee\lambda_+]_{\hbox{{\bf 1}}} \ast 1,
$$
where $\lambda_+,\lambda^{\prime}_+\in \Lambda_+$. 
It should be noted that
the above relation is satisfied only when we use the following
relation: $\hbox{Tr}(\Sigma_\alpha\Sigma_\beta)=\hbox{Tr}(\Sigma_\beta
\Sigma_\alpha)$ . This is in contrast with the generalized gauge theory of
the previous section where we needed to introduce supertrace: $\hbox{Str}
(\Sigma_\alpha\Sigma_\beta)=-\hbox{Str}(\Sigma_\beta\Sigma_\alpha)$ to
define odd-dimensional version of generalized Chern-Simons actions 
and even-dimensional topological Yang-Mills actions~\cite{KW2}.

It is important to recognize here that the gauge invariance is assured only
with the 0-form gauge parameter as is defined in (\ref{ggp zero}). As we
mentioned already in the previous section, the gauge invariance of the
action $S_G$ is lost if we introduce higher form gauge parameters due to the
breakdown of associativity by the mixture of wedge and cup products. This is
in contrast with the generalized Chern-Simons actions and the generalized
topological Yang-Mills actions where the gauge invariance with full degrees
of differential form for the gauge parameters is assured. It is important to
realize at this stage that the associativity of the generalized gauge
invariance with full degrees of differential form for the gauge parameters
will be recovered if we introduce the cup product from the beginning for the
definition of the curvature and the generalized gauge transformation. In
this case we are forced to omit exterior derivative in the definition of the
generalized curvature and the generalized gauge transformation since it
includes prohibited terms otherwise. This type of the generalized Yang-Mills
actions without derivative terms, which we shall write down in the last
section, may be related to the ``reduced model" and need further
investigations. 

Here we explicitly show the four-dimensional generalized Yang-Mills action. 
\begin{eqnarray}
S_4 = \frac{1}{e^2}\int d^Dx \sqrt{g} 
\hbox{Tr}\Big[&& ({\mathcal{F}}^{(0)})^2 + 
{\mathcal{F}}^{(1)}_{\mu}{\mathcal{F}}^{(1)\mu} - \frac{1}{2!}  
{\mathcal{F}}^{(2)}_{\mu\nu}{\mathcal{F}}^{(2)\mu\nu}  \nonumber \\
 &&-\frac{1}{3!} {\mathcal{F}}^{(3)}_{\mu\nu\rho} 
{\mathcal{F}}^{(3)\mu\nu\rho} + \frac{1}{4!}  
{\mathcal{F}}^{(4)}_{\mu\nu\rho\sigma} 
{\mathcal{F}}^{(4)\mu\nu\rho\sigma} \Big],  
\label{gga4}
\end{eqnarray}
where the explicit form of ${\mathcal{F}}^{(p)}_{\mu_1 \cdots \mu_p} \ (p=0
\sim 4)$ in terms of the generalized gauge fields including all the degrees
of differential forms $\phi,\ \omega, \ B, \ \Omega, \ H$ are given in (\ref
{defofcurv}). The explicit form of two- and three-dimensional generalized
Yang-Mills actions can be obtained similarly.

These generalized Yang-Mills actions are gauge invariant under the
generalized gauge transformation (\ref{ggt3}) or equivalently, 
\begin{eqnarray*}
&&\delta \phi= [\phi+m, v], \quad \delta \omega_\mu=\partial_\mu v +
[\omega_\mu,v],   \\
&&\delta B_{\mu\nu} = [B_{\mu\nu},v], \quad \delta \Omega_{\mu\nu\rho} =
[\Omega_{\mu\nu\rho},v] , \quad \delta H_{\mu\nu\rho\sigma} =
[H_{\mu\nu\rho\sigma},v],  
%\label{ggt4}
\end{eqnarray*}
where we have used the notation for the $p$-form gauge field: $A^{(p)}=
\frac{1}{p!}A^{(p)}_{\mu_1 \cdots \mu_p} dx^{\mu_1} \cdots dx^{\mu_p}$.

Here we explicitly give the simplest version of the four-dimensional
generalized Yang-Mills action including 0- and 1-form gauge fields only. In
this particular case ${\mathcal{F}}^{(3)}={\mathcal{F}}^{(4)}=0$ and then
the action leads much simpler form than (\ref{gga4}), 
\begin{eqnarray}
S_4 = \frac{1}{e^2}\int d^Dx \sqrt{g} 
\hbox{Tr}\Big[ &&-\frac{1}{2}F_{\mu\nu} \ F^{\mu\nu} \nonumber \\
   &&+ (\partial_\mu\phi + [\omega_\mu, \phi+m ])
       (\partial^\mu\phi + [\omega^\mu, \phi+m ]) \nonumber \\  
   &&+ (\phi+m)^4 \Big],  
\label{gga4-2}
\end{eqnarray}
where $F_{\mu\nu}=\partial_\mu\omega_\nu - \partial_\nu\omega_\mu +
[\omega_\mu,\omega_\nu]$. It is interesting to note that this action is
closely related with the Yang-Mills action $\grave{\mathit{a}}$ \textit{la}
noncommutative geometry formulation of Connes which includes only 0- and
1-form gauge fields~\cite{Connes1,Connes2,Connes3}.

%%%%%%%%%%%%%%%%%%%%

\section{Dirac-K\"ahler matter fermions}

\setcounter{equation}{0}
\setcounter{footnote}{0}

The generalized Chern-Simons actions formulated in section 2 have
topological nature because all the gauge fields and parameters are expressed
by differential forms whose metric independence are trivial from the
definitions. If we, however, try to generalize the formulation to Yang-Mills
action, the topological nature is lost because we need to introduce Hodge
star operation to define the dual of curvature, and thus need to choose a
particular metric.

As we have shown in the previous section, the generalized Yang-Mills actions
are formulated by the differential forms. In this section we formulate
matter fermions by the antisymmetric tensor of differential forms,
Dirac-K\"ahler fermion formulation. The basic idea is as follows~\cite
{DK1,DK2,DK3,BJ1,BJ2,BJ3,BJ4,BJ5,BJ6}. We first note the following well known
relations on the flat spacetime: 
\begin{equation}
(d +\delta)^2= \partial^\mu \partial_\mu = (\gamma^\mu\partial_\mu)^2,
\end{equation}
where $\delta$ is the adjoint of the exterior derivative $d$ and can be
expressed as $\delta = (-1)^{Dp +D +1} *d*$ for $p$-form in $D$-dimensional
Minkowski spacetime. In even dimensions it has the following form: 
\begin{equation}
\delta = -*d* \equiv e^\mu \partial_\mu,
\end{equation}
where we have introduced the operator $e^\mu$ which coincides with the one
defined in (\ref{cop}). The above relation suggests the
following correspondences: 
\begin{eqnarray*}
d+\delta=(dx^\mu\wedge+e^\mu)\partial_\mu &\sim& \gamma^\mu\partial_\mu, \\
dx^\mu\wedge+e^\mu &\sim& \gamma^\mu.
\end{eqnarray*}
We now reintroduce the simplest version of the cup product by 
\begin{equation}
(dx^\mu\wedge+e^\mu)\Phi \equiv dx^\mu\vee\Phi.
\end{equation}
This is the particular example of the general definition (\ref{cp}) with 
$u=dx_\mu$. We can then show that the cup product satisfies the Clifford
algebra: 
\begin{equation}
\{ dx^\mu, dx^\nu\}_{\vee}=2g^{\mu\nu},
\end{equation}
where the anticommutator with $\vee$ is defined as 
$$
\{u, v\}_\vee = u\vee v+ v\vee u,  
\label{cup product of anti-commutator}
$$
for arbitrary differential forms $u$ and $v$.

We now define the following $2^{\frac{D}{2}}\times 2^{\frac{D}{2}}$ matrix 
$Z_{ij}$ 
\begin{equation}
Z_{ij}=\sum_{p=0}^{D}\frac{1}{p!}(\gamma^T_{\mu_1}\cdot\cdot\cdot
\gamma^T_{\mu_p})_{ij}dx^{\mu_1}\wedge\cdot\cdot\cdot\wedge dx^{\mu_p}.
\end{equation}
which satisfies the following authogonality: 
\begin{equation}
\int Z_{ij}\vee Z_{kl} * 1 = 2^{\frac{D}{2}} \delta_{il}\delta_{jk}\int *1.
\label{orthogonality of Z}
\end{equation}
We can then show the following crucial relation: 
\begin{equation}
dx^\mu\vee Z_{ij}=(\gamma^{\mu T} Z)_{ij}=\gamma^\mu_{ki}Z_{kj}.
\label{crucial relation}
\end{equation}

We now consider an arbitrary fermionic field $\Psi$ which is a direct sum of
fermionic antisymmetric tensor fields contracted with differential forms.
Then this form valued generalized fermionic field $\Psi$ can be expanded by 
$Z_{ij}$, 
\begin{equation}
\Psi(x)=\sum_{i,j}\psi_{ij}(x)Z_{ij},
\end{equation}
where 
\begin{equation}
\psi_{ij}(x) = 2^{-\frac{D}{2}}[\Psi(x)\vee Z_{ji}]_0,
\label{psi expansion coefficient}
\end{equation}
where $[\quad]_0$ denotes to pick up 0-form term. By employing the relation 
(\ref{crucial relation}), we obtain 
\begin{equation}
d\vee \Psi = (\gamma^\mu)_{ki}\partial_\mu \psi_{ij}Z_{kj}.
\end{equation}

Considering the completeness of $Z_{ij}$ to be understood from the
orthogonality relation (\ref{orthogonality of Z}), we obtain the following
Dirac-K\"ahler equations which have $2^{\frac{D}{2}}$ ``flavor"
component suffices $i$: 
\begin{equation}
d\vee \Psi = 0 \longleftrightarrow (\gamma^\mu)_{\alpha\beta} \partial_\mu
\psi_{\beta (i)} = 0,
\end{equation}
where the first suffix of $\psi_{ij}$ is identified as spinor suffix. We
then obtain the free Dirac fermion Lagrangian from the Dirac-K\"ahler
fermion formulation, 
\begin{equation}
\int \overline{\Psi}\vee d\vee\Psi = 2^{\frac{D}{2}}\sum_{j}\int d^Dx 
\overline{\psi}_{(j)}\partial\llap{/} \psi_{(j)},
\end{equation}
where we obtain $2^{\frac{D}{2}}$ copy of flavor suffices $(j)$.

We now introduce generalized version of our fermionic matter fields by
Dirac-K\"ahler fermion formulation. We first define a direct sum of the
quaternion valued fermionic antisymmetric tensor fields contracted with
differential forms: 
\begin{equation}
\begin{array}{rcl}
{\Psi} &=& \mbox{\bf{1}}\psi + \mbox{\bf{i}}\hat{\psi} \quad\in \Lambda_-, \\
\overline{\Psi} &=& \mbox{\bf{j}}\hat{\overline{\psi}} + \mbox{\bf{k}}
\overline{\psi} \quad\in \Lambda_+,  
\end{array}
\label{gff}
\end{equation}
where $\psi$ and $\overline{\psi}$ are fermionic odd forms while $\hat{\psi}$
and $\hat{\overline{\psi}}$ are fermionic even forms. The conjugate
fermionic field $\overline{\Psi}$ is defined to be related with ${\Psi}$ as: 
\begin{eqnarray}
\overline{\Psi} &=& \zeta(\Psi^\dagger \vee \mbox{\bf{j}}dx^0) \quad\in
\Lambda_+ \ (\mbox{Minkowski}),  
\label{minkowski conjugate} \\
\overline{\Psi} &=& \zeta(\Psi^\dagger \mbox{\bf{k}}) \quad\in \Lambda_+ \ 
(\mbox{Euclid}),  
\label{euclidean conjugate}
\end{eqnarray}
where we define 
${\Psi}^\dagger = ( \mbox{\bf{1}}\psi + \mbox{\bf{i}}\hat{\psi})^\dagger
= \mbox{\bf{1}}\psi^\dagger + \mbox{\bf{i}}\hat{\psi}^\dagger$.  
The field $\overline{\Psi}$ has different $\Psi^\dagger $ dependence from
the Minkowskian case to the Euclidean case analogous to the usual
expressions. They, however, belong to the same $\Lambda_+$ class. We then
obtain the following relations: 
\begin{eqnarray}
\hat{\overline{\psi}} &=& \zeta(\psi^\dagger\vee dx^0), 
\quad \overline{\psi} = \zeta(\hat{\psi}^\dagger\vee dx^0) 
\quad(\mbox{Minkowski}), \\
\hat{\overline{\psi}} &=& \epsilon_1\zeta\hat{\psi}^\dagger, \hspace{15mm}
\overline{\psi} = \zeta\psi^\dagger \quad (\mbox{Euclid}).  
%\label{gff2}
\end{eqnarray}

With all these definitions given we introduce generalized matter action
coupled to the generalized gauge fields 
\begin{equation}
S_M = \int \Big[\overline{\Psi} \vee 
                (\mathcal{D+A})\vee\Psi\Big]_{\mbox{\bf{1}}}
* 1,  
\label{gmaction}
\end{equation}
where the generalized differential operator $\mathcal{D}$ and the
generalized gauge field $\mathcal{A}$ are given in (\ref{gdo2}) and 
(\ref{ggf2}), respectively. The reason of taking the coefficient of 
${\mbox{\bf{1}}}$ for the quaternion expansion in the trace is the same as 
the generalized Yang-Mills action. 
In other words we need to pick up 0-form components in
the trace and thus take the even form part of ${\mbox{\bf{1}}}$-th component
due to the presence of $* 1$.
In this paper we consider that fermions belong to the fundamental 
representation of the gauge algebra.

Substituting the generalized fermionic fields (\ref{gff}) into the
generalized matter action $S_M$ and collecting the ${\mbox{\bf{1}}}$-th 
components, we obtain 
\begin{eqnarray}
S_M = \epsilon_2\int \Big[&&(\epsilon_1\overline{\psi}
+ \hat{\overline{\psi}})\vee (d+A)\vee (\psi+\hat{\psi}) \nonumber \\
&& - \epsilon_1(\overline{\psi}+\hat{\overline{\psi}})\vee (\hat{A} +m)\vee
(\psi+\hat{\psi}) \Big]*1,  
\label{generalized matter action}
\end{eqnarray}
where $\epsilon_1$ and $\epsilon_2$ are the sign factors of quaternion
algebra (\ref{quaternion}). 
It should be noted that only the 0-form terms
are allowed in the square bracket of the generalized matter action $S_M$ due
to the presence of $*1$. There are thus several trivial terms included in
the matter action such as $\overline{\psi}\vee (d+A)\vee \psi *1 =0$. 
In order that $\overline{\Psi}=\overline{\psi}+\hat{\overline{\psi}}$ be the
common conjugate fermionic field in the generalized matter action (\ref
{generalized matter action}), we need to take the particular choice of the
sign factor $\epsilon_1=+1$ for the quaternion algebra (\ref{quaternion})
while the overall sign factor $\epsilon_2$ could be arbitrary and thus 
$\epsilon_2=\pm 1$.

There are two possible quaternion algebra which are consistent with the
conjugate definitions of the generalized fermionic fields (\ref{minkowski
conjugate}) and (\ref{euclidean conjugate}). 
The first ``quaternion algebra" corresponds to the 
choice $(\epsilon_1,\epsilon_2)=(+1,+1)$: 
\begin{equation}
(1) \quad {\mbox{\bf{i}}}^2={\mbox{\bf{j}}}^2={\mbox{\bf{1}}}, \quad 
{\mbox{\bf{k}}}^2= -{\mbox{\bf{1}}}, \quad 
{\mbox{\bf{i}}}{\mbox{\bf{j}}}={\mbox{\bf{k}}}, \quad 
{\mbox{\bf{j}}}{\mbox{\bf{k}}}=-{\mbox{\bf{i}}}, \quad 
{\mbox{\bf{k}}}{\mbox{\bf{i}}}=-{\mbox{\bf{j}}},  
\label{quaternion algebra 1}
\end{equation}
while the second ``quaternion algebra" corresponds to the choice $
(\epsilon_1, \epsilon_2) = (+1,-1)$: 
\begin{equation}
(2) \quad {\mbox{\bf{i}}}^2={\mbox{\bf{k}}}^2={\mbox{\bf{1}}}, \quad 
{\mbox{\bf{j}}}^2= -{\mbox{\bf{1}}}, \quad 
{\mbox{\bf{i}}}{\mbox{\bf{j}}}={\mbox{\bf{k}}}, \quad 
{\mbox{\bf{j}}}{\mbox{\bf{k}}}={\mbox{\bf{i}}}, \quad 
{\mbox{\bf{k}}}{\mbox{\bf{i}}}=-{\mbox{\bf{j}}}.  
\label{quaternion algebra 2}
\end{equation}
These algebra are common to Minkowski and Euclidean cases.

Hereafter we take the choice of the first ``quaternion algebra" (\ref
{quaternion algebra 1}) and then the generalized matter action leads 
\begin{eqnarray}
S_M &=& i2^{-\frac{D}{2}}\int\Big[\overline{\Psi} \vee (\mathcal{D + A})
\vee\Psi\Big]_{\mbox{\bf{1}}}* 1  \nonumber \\
&=& i2^{-\frac{D}{2}}\int\Big[(\overline{\psi}+\hat{\overline{\psi}})\vee
(d+A - \hat{A} -m)\vee (\psi+\hat{\psi})\Big]*1,
\label{generalized matter action 1}
\end{eqnarray}
where we have introduced the normalization factor.

Let us now define 
\begin{eqnarray}
\psi + \hat{\psi} &=& \psi_{ij}Z_{ij},  \nonumber \\
\overline{\psi} + \hat{\overline{\psi}} &=& \overline{\psi}_{ij}Z_{ij},  \\
A-\hat{A} &=&-\sum^D_{p=0} \frac{(-1)^p}{p!} A^{(p)}_{\mu_1 \cdots \mu_p}
dx^{\mu_1}\wedge \cdots \wedge dx^{\mu_p}.  \nonumber
\label{new definitions of gauge and matter fields}
\end{eqnarray}
Using the relation $dx^\mu\wedge dx^\nu=dx^\mu\vee dx^\nu-g^{\mu\nu}$
successively and the relation (\ref{crucial relation}), we can prove the
following relation: 
\begin{eqnarray*}
&&A^{(p)}_{\mu_1 \cdots \mu_p} (dx^{\mu_1}\wedge \cdots \wedge dx^{\mu_p})
\vee \psi_{ij}Z_{ij} \\
&=& A^{(p)}_{\mu_1 \cdots \mu_p} (dx^{\mu_1}\vee \cdots
\vee dx^{\mu_p}) \vee \psi_{ij}Z_{ij}  \\
&=& A^{(p)}_{\mu_1 \cdots \mu_p} (\gamma^{\mu_1} \cdots
\gamma^{\mu_p})_{ki}\psi_{kj}Z_{ij}.  
%\label{new relation}
\end{eqnarray*}
We can then obtain the following concrete expression of the generalized
matter action coupled to the generalized gauge fields: 
\begin{equation}
S_M = \int d^Dx
\Big[\overline{\psi}_{ik}i(\gamma^\mu\partial_\mu - \sum^D_{p=0}
\frac{(-1)^p}{p!}A^{(p)}_{\mu_1 \cdots \mu_p} (\gamma^{\mu_1} \cdots
\gamma^{\mu_p})-m)_{kl}\psi_{li}\Big].
\label{concrete expression of the generalized matter action }
\end{equation}

We next consider the gauge invariance of the generalized matter action $S_M$. 
It is easy to show that $S_M$ is gauge invariant if the generalized gauge
fields and matter fields transform as: 
\begin{eqnarray}
&&\delta {\mathcal{A}} = [{\mathcal{D}} + {\mathcal{A}}, {\mathcal{V}}]_\vee,
\label{ggt6} \\
&&\delta \Psi = - {\mathcal{V}}\vee \Psi, \quad  
\delta \overline{\Psi} = \overline{\Psi}\vee {\mathcal{V}},  
\label{ggt5} 
\end{eqnarray}
where the gauge transformation of the generalized gauge field $\mathcal{A}$
has the similar transformation form as the generalized gauge transformation 
(\ref{ggt2}) except that the commutator with cup product 
%(\ref{cup product of commutator}) 
is adapted. It is important to realize here again that the
generalized gauge transformation of the gauge field (\ref{ggt6}) includes
prohibited terms unless we omit exterior derivative. This situation is
similar to the generalized Yang-Mills action as we have mentioned in the
previous section. If we, however, introduce only 0-form gauge parameter as
in (\ref{ggp zero}), the gauge invariance with differential operator is
recovered even with full degrees of differential forms for the generalized
gauge fields and matter fermions.

Since $\overline{\Psi}$ is related to $\Psi$ by the relations (\ref
{minkowski conjugate}) for Minkowski case and (\ref{euclidean conjugate})
for Euclidean case, it is not obvious if the generalized gauge
transformation (\ref{ggt5}) is consistent. As far as the generalized gauge
parameter includes only 0-form, we can consistently impose the gauge
transformation (\ref{ggt5}) by taking anti-Hermite 0-form gauge parameter.

Here we investigate the consistency condition between the generalized gauge
transformation (\ref{ggt5}) and the definition of the conjugate fermionic
field (\ref{minkowski conjugate}) in the Minkowski case. We study the
general case for the generalized gauge parameter including all the degrees
of differential forms 
\begin{equation}
\mathcal{V} = {\mbox{\bf 1}} \hat{a} + {\mbox{\bf i}}a,  
\label{ggp2}
\end{equation}
where $\hat{a}$ and $a$ are the direct sum of bosonic even and odd forms,
respectively. For simplicity we have omitted the fermionic gauge parameters
here, to be compared with the general expression (\ref{ggp0}).

Due to the relation between the conjugate fermionic field $\overline{{\Psi}}$
and the original fermionic field ${\Psi}$ given 
in (\ref{minkowski conjugate}), the gauge transformation 
of the conjugate fermionic field should obey 
\begin{eqnarray}
\delta \overline{\Psi} &=& \zeta(\delta\Psi^\dagger \vee \mbox{\bf{j}}dx^0)
\nonumber \\
&=& \mbox{\bf{j}}(-\hat{\overline{\psi}}\vee \zeta \hat{a}^\dagger - 
\overline{\psi}\vee \zeta a^\dagger) + \mbox{\bf{k}}(\overline{\psi}\vee
\zeta \hat{a}^\dagger - \hat{\overline{\psi}}\vee \zeta a^\dagger),
\end{eqnarray}
while $\delta \overline{\Psi}$ needs to satisfy the gauge transformation 
(\ref{ggt5}), 
\begin{eqnarray}
\delta \overline{\Psi} &=& \overline{\Psi}\vee \mathcal{V}  \nonumber \\
&=& \mbox{\bf{j}}(\hat{\overline{\psi}}\vee \hat{a} - \overline{\psi}\vee a)
+ \mbox{\bf{k}}(-\overline{\psi}\vee \hat{a} - \hat{\overline{\psi}}\vee a).
\end{eqnarray}
To be consistent, these two expressions should coincide and thus we obtain
the following consistency constraints: 
\begin{equation}
\zeta \hat{a}^\dagger = -\hat{a}, \quad \zeta a^\dagger = a.
\label{consistency condition}
\end{equation}
It turns out that these constraints for the Euclidean case exactly coincide
with those of the Minkowski case.

In these derivations we have used the sign operator property (\ref
{sfrelations}) and adapted the following Hermite conjugate: 
$$
\zeta (\hat{a} \vee \psi)^\dagger \ =\ \zeta(\hat{a}^{b*}\vee \psi^\dagger)
T_b^\dagger \ = \ (\zeta\psi^\dagger)\vee(\zeta \hat{a}^{b*})T_b^\dagger. 
$$
%

%%%%%%%%%%%%%%%%%%%%

\section{Weinberg-Salam model from generalized gauge theory}

\setcounter{equation}{0}
\setcounter{footnote}{0}

In formulating the generalized Yang-Mills actions with Dirac-K\"ahler
fermions, we have not specified the gauge algebra. In formulating the
generalized Chern-Simons actions and topological Yang-Mills actions, the
graded Lie algebra as a gauge algebra of the generalized gauge theory was
the natural consequence of the formulation. This characteristic of the
natural introduction of the graded Lie algebra in the generalized gauge
theory transfers to the formulation of the generalized Yang-Mills actions.
In this section we take a particular graded Lie algebra of supergroup 
$SU(2|1)$, as the generalized gauge algebra and show that the Weinberg-Salam
model with spontaneously broken symmetry can be formulated naturally from
the generalized Yang-Mills action. Though the noncommutative geometry
formulation of the Weinberg-Salam model based on the $SU(2|1)$ graded Lie
algebra was intensively studied by Coquereaux et al.~\cite{Coq4}, we believe
that our formulation based on the generalized gauge theory will give new
insights into the formulation.

In order to accommodate the Dirac-K\"ahler fermion formulation into our
generalized gauge theory formulation we needed to specify the quaternion
algebra in such a way that the generalized gauge transformation of the
fermion and the conjugate definitions of the generalized fermionic field are
consistent. Here in this section we choose the following ``quaternion
algebra" defined in (\ref{quaternion algebra 1}) 
\begin{eqnarray}
(1) \quad {\mbox{\bf{i}}}^2={\mbox{\bf{j}}}^2={\mbox{\bf{1}}}, \quad 
{\mbox{\bf{k}}}^2= -{\mbox{\bf{1}}}, \quad 
{\mbox{\bf{i}}}{\mbox{\bf{j}}}={\mbox{\bf{k}}}, \quad 
{\mbox{\bf{j}}}{\mbox{\bf{k}}}=-{\mbox{\bf{i}}}, \quad 
{\mbox{\bf{k}}}{\mbox{\bf{i}}}=-{\mbox{\bf{j}}}.  
\label{quaternion algebra 3}
\end{eqnarray}

We now introduce the algebra of supergroup $SU(2|1)$ as the graded Lie
algebra. $SU(2|1)$ generators can be represented by $3\times3$ matrices~\cite
{Rittenberg1,Rittenberg2,Rittenberg3,Coq4} 
%
%%%%%
\renewcommand{\arraystretch}{1.0}
%%%%%
%%% new array
\makeatletter
\def\@arrayacol{\edef\@preamble{\@preamble \hskip .7\arraycolsep}}
\def\array{\let\@acol\@arrayacol \let\@classz\@arrayclassz
\let\@classiv\@arrayclassiv \let\\\@arraycr\def\@halignto{}\@tabarray}
\makeatother
$$
T_{i}=\left( 
\begin{array}{@{\,}cc|c}
& 
\mbox{ \hspace{-7.6mm} \raisebox{-2.5mm}[0pt][0pt]
                  { $ \displaystyle \ \frac{\sigma_{i}}{2} $ } } & 0 \\ 
&  & 0 \\ \hline
0 & 0 & 0
\end{array}
\right), \quad  
Y =\left( 
\begin{array}{@{\,}cc|c}
1 & 0 & 0 \\ 
0 & 1 & 0 \\ \hline
0 & 0 & 2
\end{array}
\right), 
\]
\[
\Sigma_{+}=\left( 
\begin{array}{@{\,}cc|c}
0 & 0 & 1 \\ 
0 & 0 & 0 \\ \hline
0 & 0 & 0
\end{array}
\right), \quad  
\Sigma_{-}=\left( 
\begin{array}{@{\,}cc|c}
0 & 0 & 0 \\ 
0 & 0 & 1 \\ \hline
0 & 0 & 0
\end{array}
\right), \quad  
\Sigma^{\prime}_{+}=\left( 
\begin{array}{@{\,}cc|c}
0 & 0 & 0 \\ 
0 & 0 & 0 \\ \hline
0 & 1 & 0
\end{array}
\right), \quad  
\Sigma^{\prime}_{-}=\left( 
\begin{array}{@{\,}cc|c}
0 & 0 & 0 \\ 
0 & 0 & 0 \\ \hline
1 & 0 & 0
\end{array}
\right), \ 
$$
%
%%% new array
\makeatletter
\def\@arrayacol{\edef\@preamble{\@preamble \hskip .2\arraycolsep}}
\def\array{\let\@acol\@arrayacol \let\@classz\@arrayclassz
\let\@classiv\@arrayclassiv \let\\\@arraycr\def\@halignto{}\@tabarray}
\makeatother
where $\sigma_{i}$'s are Pauli matrices. They satisfy the following graded
Lie algebra: 
\[
[ T_{i}, T_{j} ]=i\epsilon_{ijk}T_{k}, \quad  
[ Y, T_{i} ]=0, 
\]
\[
[ T_{\pm}, \Sigma_{\pm} ]=0, \quad  
[ T_{\pm}, \Sigma_{\mp} ]=\frac{1}{\sqrt{2}}\Sigma_{\pm}, 
\]
\[
[ T_{\pm}, \Sigma^{\prime}_{\pm} ]=0, \quad  
[ T_{\pm}, \Sigma^{\prime}_{\mp} ]=-\frac{1}{\sqrt{2}}\Sigma^{\prime}_{\pm}, 
\]
\[
[ T_{3}, \Sigma_{\pm} ]=\pm\frac{1}{2}\Sigma_{\pm}, \quad  
[ T_{3}, \Sigma^{\prime}_{\pm} ]=\pm\frac{1}{2}\Sigma^{\prime}_{\pm}, 
\]
\[
[ Y, \Sigma_{\pm} ]=-\Sigma_{\pm}, \quad  
[ Y, \Sigma^{\prime}_{\pm} ]=\Sigma^{\prime}_{\pm}, 
\]
\[
\{ \Sigma_{\pm}, \Sigma_{\pm} \}  = \{ \Sigma_{\pm}, \Sigma_{\mp} \}=0, \quad
\{ \Sigma^{\prime}_{\pm}, \Sigma^{\prime}_{\pm} \} =
\{ \Sigma^{\prime}_{\pm}, \Sigma^{\prime}_{\mp} \}=0, 
\]
\[
\{ \Sigma_{\pm}, \Sigma^{\prime}_{\pm} \}=\sqrt{2}T_{\pm}, \quad  
\{ \Sigma_{\pm}, \Sigma^{\prime}_{\mp} \}=\pm T_{3}+\frac{1}{2}Y, 
\]
with $T_{\pm}=\frac{1}{\sqrt{2}}(T_{1}\pm iT_{2})$. $T_{3}$, $Y$ correspond to
a generator of a weak isospin and a weak hypercharge, respectively. This
algebra contains $SU(2)\times U(1)_{Y}$ in the even graded parts $T_{i}$, $Y$,
and $SU(2)$ doublets in the odd graded parts $\Sigma_{\pm}$, 
$\Sigma^{\prime}_{\pm}$ whose subscripts $\pm$ correspond to the generators
with eigenvalues $\pm\frac{1}{2}$ of the generator $T_{3}$. Indeed Higgs
doublet which we have introduced as Lie algebra valued gauge fields
corresponds to these odd parts. It is interesting to note that the 
supersymmetric algebra of $SU(2|1)$ can be accommodated in the generalized 
gauge theory even without fermionic fields.

In the formulation of generalized Yang-Mills action of the previous section
we have introduced all the degrees of differential forms but only 0-form
for the generalized gauge parameter. The reason why we have introduced only
0-form gauge parameter is that the action is not gauge invariant under
the higher form gauge parameters. Hereafter we introduce only 0-form and
1-form gauge fields in accordance with the standard gauge theory.

The gauge field is now expanded by corresponding fields of the generators 
\begin{equation}
{\mathcal{A}}={\mbox{\bf{j}}} ie\Big(A^{i}T_{i}+\frac{1}{2\sqrt{3}}BY\Big) 
+{\mbox{\bf{k}}} \frac{ie}{\sqrt{2}}\Big(\phi_{0}\Sigma_{+}+\phi_{+}\Sigma_{-}
+\phi_{0}^{*}\Sigma^{\prime}_{-} +\phi_{+}^{*}\Sigma^{\prime}_{+}\Big),
\label{weinberg-salam gauge field}
\end{equation}
where $A^{i}$, $B$ and $\phi_{0}$, $\phi_{+}$ are real $SU(2)\times U(1)_{Y}$
1-form gauge fields and 0-form complex Higgs scalar fields respectively. The
normalization factors and the pure imaginary constant $i$ of each fields are
adjusted to give the standard kinetic terms in the final Weinberg-Salam
action. The generalized gauge field is rescaled by the coupling constant $e$: 
${\mathcal{A}}\rightarrow e{\mathcal{A}}$.

We can choose a particular form of the constant matrix $m$ of the
generalized differential operator (\ref{gdo2}) parametrized by a complex 
number $v$ as 
\begin{eqnarray}
\mathcal{D} &=& \mbox{\bf{j}}d + \mbox{\bf{k}}m^{\alpha}\Sigma_{\alpha} 
\nonumber \\
&=& \mbox{\bf{j}}d + \mbox{\bf{k}}\frac{i}{\sqrt{2}}(v\Sigma_{+}
+v^*\Sigma^{\prime}_{-}),  
\label{gdo3}
\end{eqnarray}
which leads 
\begin{equation}
\mathcal{D}^2= - {\mbox{\bf{1}}}m^2 =
{\mbox{\bf{1}}}\frac{|v|^2}{2}\Big(T_{3}+\frac{1}{2}Y\Big).  
\label{em}
\end{equation}
Then $\mathcal{D}^2$ can be taken to be proportional to the generator of the
electromagnetic charge of the Weinberg-Salam model.

The generalized curvature is now given by 
\begin{eqnarray}
\mathcal{F} &\equiv& \mathcal{D}^2 + \{ \mathcal{D}, \mathcal{A} \} + 
\mathcal{A}^2  \nonumber \\
&=& {\mbox{\bf{1}}} \Big(e^2{\mathcal{F}}^{(0)} 
-\frac{i}{2}e{\mathcal{F}}_{\mu\nu}^{(2)} dx^{\mu}\wedge dx^{\nu} \Big) 
+{\mbox{\bf{i}}}ie{\mathcal{F}}_{\mu}^{(1)}dx^{\mu}.
\end{eqnarray}
The kinetic terms of $SU(2)\times U(1)_{Y}$ gauge fields are 
\begin{equation}
{\mathcal{F}}_{\mu\nu}^{(2)}=F_{\mu\nu}^kT_{k}+\frac{1}{2\sqrt{3}}G_{\mu\nu}Y,
\end{equation}
where 
\begin{eqnarray*}
&&F_{\mu\nu}^{k} = \partial_{\mu}A_{\nu}^{k}-\partial_{\nu}A_{\mu}^{k} 
-e\epsilon_{ij}{}^{k}A_{\mu}^{i}A_{\nu}^{j}, \\
&&G_{\mu\nu}=\partial_{\mu}B_{\nu}-\partial_{\nu}B_{\mu}. 
\end{eqnarray*}
The kinetic terms of Higgs fields and the gauge-Higgs interaction terms are 
\begin{eqnarray}
{\mathcal{F}}_{\mu}^{(1)}=
-\frac{1}{\sqrt{2}}\Bigg\{ &\bigg(&\partial_{\mu}\phi_{0} 
+\frac{i}{\sqrt{2}}eW_{\mu}\phi_{+} +\frac{i}{\sqrt{3}}eZ_{\mu} 
\Big(\phi_{0}+\frac{v}{e}\Big) \bigg)\Sigma_{+}  \nonumber \\
+ &\bigg(&\partial_{\mu}\phi_{+} 
+\frac{i}{\sqrt{2}}eW_{\mu}^{\dagger} \Big(\phi_{0}+\frac{v}{e}\Big) 
-i\frac{\sqrt{3}}{6}eZ_{\mu}\phi_{+} -\frac{i}{2}eA_{\mu}\phi_{+} 
\bigg)\Sigma_{-}  \nonumber \\
+&{}&{\mbox{h.c.}} \ \Bigg\},
\end{eqnarray}
where
\[
W_{\mu}=\frac{1}{\sqrt{2}}\Big( A_{\mu}^{1}-iA_{\mu}^{2} \Big), 
\]
and
% 
%%%%%
\renewcommand{\arraystretch}{2.0}
%%%%%
%
\begin{equation}
\begin{array}{rcl}
Z_{\mu} &=& \displaystyle\frac{\sqrt{3}}{2}A_{\mu}^3-\frac{1}{2}B_{\mu},   \\
A_{\mu} &=& \displaystyle\frac{1}{2}A_{\mu}^3+\frac{\sqrt{3}}{2}B_{\mu}.  
\end{array}
\label{az}
\end{equation}
These identifications (\ref{az}) fix the Weinberg angle to be 
$\theta_{W}=\frac{\pi}{6}$ which is an arbitrary parameter 
in the Weinberg-Salam model.
Thus the direction of spontaneous breaking is particularly chosen by the
model itself. The Higgs potential term is given by 
\begin{eqnarray}
{\mathcal{F}}^{(0)}=\frac{1}{2} &&\Bigg\{\Big|\phi_{0}+\frac{v}{e}\Big|^2 
\Big(T_{3}+\frac{1}{2}Y\Big) +\bigg(\phi_{0}+\frac{v}{e}\bigg) 
\phi_{+}^{*}\sqrt{2}T_{+}  \nonumber \\
&&+\phi_{+}\bigg(\phi_{0}^{*}+\frac{v^{*}}{e}\bigg) \sqrt{2}T_{-}
+|\phi_{+}|^2\Big(-T_{3}+\frac{1}{2}Y\Big)\Bigg\}.  \label{c0}
\end{eqnarray}

As we have seen before, the generalized gauge transformation of the
generalized curvature is given by (\ref{gtgc1}): 
$\delta\mathcal{F}=[\mathcal{F}, \mathcal{V}_0]_\vee$. 
The 0-form generalized gauge parameter 
${\ \mathcal{V}}_0= \hbox{{\bf 1}} v^aT_a$ depends only on the even part 
of generators of
the graded $SU(2|1)$ algebra and thus commutes with the generator $Y$.
Therefore the gauge transformation of the generalized curvature is form
invariant even if we add the term proportional to the generator $Y$ to the
curvature. In other words there is a particular arbitrariness of the
constant term in the definition of the generalized curvature. We thus define
a new curvature which includes the term,
\begin{equation}
{\mathcal{F}}^{\prime}= {\mathcal{F}} + y|v|^2 Y,
\label{new definition of curvature}
\end{equation}
where we introduce a new parameter $y$ with previously introduced 
dimensionful parameter $v$ in (\ref{gdo3}).

Using the new definition of the generalized curvature, we obtain the full
expression of the generalized Yang-Mills action with $SU(2|1)$ graded Lie
algebra in flat Minkowski spacetime
\begin{eqnarray}
S_G &=& -\frac{1}{e^2} \int \hbox{Tr}\Big[{\mathcal{F}}^{\prime}\vee 
{\mathcal{F}}^{\prime}\Big]_{\hbox{\bf 1}}*1  \nonumber \\
&=& - \int d^4x \hbox{Tr}\Big[\frac{1}{2} 
{\mathcal{F}}^{(2)}_{\mu\nu}{\mathcal{F}}^{(2)\mu\nu} 
- {\mathcal{F}}^{(1)}_{\mu}{\mathcal{F}}^{(1)\mu} +
e^2 ({\mathcal{F}}^{(0)})^2 \Big]  \nonumber \\
&=& \int d^4x \Bigg\{ - \frac{1}{4}F_{\mu\nu}F^{\mu\nu} 
- \frac{1}{4}Z_{\mu\nu}Z^{\mu\nu}  \nonumber \\
&&\hspace{14mm}
-\frac{1}{2}(D^{\mu\dagger}W^{\nu\dagger}-D^{\nu\dagger}W^{\mu\dagger})
(D_{\mu}W_{\nu}-D_{\nu}W_{\mu})  \nonumber \\
&&\hspace{14mm}
-ie\Big(\frac{\sqrt{3}}{2}Z_{\mu\nu} 
+\frac{1}{2}F_{\mu\nu}\Big)W^{\mu}W^{\nu\dagger}  \nonumber \\
&&\hspace{14mm}
+\frac{e^2}{2}\Big(|W_{\mu}W^{\mu}|^2-(W_{\mu}W^{\mu\dagger})^2\Big) 
\nonumber \\
&&\hspace{14mm}
+\Big|{\partial}_{\mu}\phi_{0}+\frac{i}{\sqrt{2}}eW_{\mu}\phi_{+} 
+\frac{i}{\sqrt{3}}eZ_{\mu} \Big(\phi_{0}+\frac{v}{e}\Big)\Big|^2  \nonumber \\
&&\hspace{14mm}
+\Big|{\partial}_{\mu}\phi_{+} +\frac{i}{\sqrt{2}}eW_{\mu}^{\dagger} 
\Big(\phi_{0}+\frac{v}{e}\Big) 
-i\frac{\sqrt{3}}{6}eZ_{\mu}\phi_{+}^* -\frac{i}{2}eA_{\mu}\phi_{+}^* 
\Big|^2  \nonumber \\
&&\hspace{14mm}
-\frac{e^2}{2}\bigg( \Big|\phi_{0}+\frac{v}{e}\Big|^2 
+ |\phi_{+}|^2 \bigg)^2  \nonumber \\
&&\hspace{14mm}
-3ye^2\Big|\frac{v}{e}\Big|^2 \bigg( \Big|\phi_{0}+\frac{v}{e}\Big|^2
+|\phi_{+}|^2 \bigg)  \nonumber \\
&&\hspace{14mm}
-6y^2e^2\Big|\frac{v}{e}\Big|^2 \Bigg\},  
\label{weinberg-salam action}
\end{eqnarray}
where
\begin{eqnarray*}
F_{\mu\nu} &=& \partial_{\mu}A_{\nu}-\partial_{\nu}A_{\mu}, \\
Z_{\mu\nu} &=& \partial_{\mu}Z_{\nu}-\partial_{\nu}Z_{\mu}, \\
D_{\mu} &=& \partial_{\mu}+ie\Big(\frac{\sqrt{3}}{2}Z_{\mu} 
+\frac{1}{2}A_{\mu} \Big).
\end{eqnarray*}
This is the Weinberg-Salam model with the Weinberg angle 
$\theta_{W}=\frac{\pi}{6}$. If we look at the Higgs potential term, 
it has the local minimum at \\ 
(1)  $|\phi_{0}+\frac{v}{e}|=0, \quad |\phi_+|=0$ \quad for $y\ge 0$, \\
(2)  $|\phi_{0}+\frac{v}{e}|=\sqrt{3|y|}\frac{|v|}{e}, \quad
   |\phi_+|=0$ \quad for $y<0$. \\ 
Then the masses of the weak bosons $W^{\pm}$, $Z$ and Higgs are,
respectively, given by \\
(1)  $M_W=0$,  $M_Z=0$,  $M_\phi=\sqrt{3|y|}|v|$, \\
(2)  $M_W=\sqrt{\frac{3|y|}{2}}|v|$,  $M_Z=\sqrt{2|y|}|v|$,  
$M_\phi=\sqrt{6|y|}|v|$, and thus $\frac{M_\phi}{M_W}=2$. \\
The Higgs-weak boson mass ratio coincides with the result of noncommutative
geometry formulation $\grave{\mathit{a}}$ \textit{la} Connes~\cite
{Connes1,Connes2,Connes3}\cite{Coq4}. It is important to recognize here that
the spontaneously broken phase can be realized only when $y<0$.

Phenomenologically the Weinberg angle $\hbox{sin}^2\theta_W = 0.25$ is close
to the experimental value $\approx 0.23$, while the Higgs-weak boson mass
ratio might be away from the recent informal result of LEP which is very
close to $\sqrt{2}$~\cite{Higgs}. It is, however, interesting to note that
0-, 1- and 2-form curvature terms, $\hbox{Tr}[({\mathcal{F}}^{(0)})^2]$, 
$\hbox{Tr}[{\mathcal{F}}^{(1)}_{\mu}{\mathcal{F}}^{(1)\mu}]$ and 
$\hbox{Tr}[{\mathcal{F}}^{(2)}_{\mu\nu}{\mathcal{F}}^{(2)\mu\nu}]$ 
in the action (\ref{weinberg-salam action}) are independently gauge 
invariant as we pointed out
in (\ref{gtgc2}). As far as the gauge invariance of the even generators of
the graded Lie algebra $SU(2|1)$ is concerned, there thus come in two free
parameters in the action except for the overall normalization factor. If we
introduce these parameters, the Weinberg angle is unchanged but the
Higgs-weak boson mass ratio will get the parameter dependence. Thus the
above predicted mass ratio, $\frac{M_\phi}{M_W}$, may be changed if we
introduce these free parameters

We now consider the generalized matter action (\ref{generalized matter
action 1}) with the generalized gauge field (\ref{weinberg-salam gauge field})
including only 0- and 1-form gauge fields, 
\begin{eqnarray}
S_M &=& \frac{i}{4}\int\Big[\overline{\Psi} \vee (\mathcal{D + A})
\vee\Psi\Big]_{\mbox{\bf{1}}}* 1  \nonumber \\
&=& \frac{i}{4}\int \Big[(\overline{\psi}+\hat{\overline{\psi}})\vee (d+A
- \hat{A} -m)\vee (\psi+\hat{\psi})\Big] * 1  \nonumber \\
&=& \int d^4x\overline{\psi}^{(j)}
    \Big\{\gamma^\mu(i\partial_\mu -e(A^i_\mu T_i 
           + \frac{1}{2\sqrt{3}}B_\mu Y))  \nonumber \\
&&\hspace{20mm} +\frac{e}{\sqrt{2}}\Big((\phi_0+\frac{v}{e})\Sigma_+ 
                + \phi_+\Sigma_- 
                + (\phi_0^* + \frac{v^*}{e})\Sigma_-^{\prime} 
                + \phi_+^*\Sigma_+^{\prime}\Big) \Big\}\psi^{(j)},
\label{weinberg-salam matter action}
\end{eqnarray}
where the following relations should be understood: 
\begin{eqnarray*}
A^{ab}&=&ie(A^i_\mu T_i + \frac{1}{2\sqrt{3}}B_\mu Y)^{ab},  \\
\hat{A}^{ab}&=&\frac{ie}{\sqrt{2}}(\phi_0\Sigma_+ +
\phi_+\Sigma_- + \phi_0^*\Sigma_-^{\prime}+
\phi_+^*\Sigma_+^{\prime})^{ab},  \\
m^{ab}&=&\frac{i}{\sqrt{2}}(v\Sigma_{+}+v^*\Sigma^{\prime}_{-})^{ab}, \\
\psi^{(j)a}_\alpha &=& \frac{1}{4}\Big[(\psi+ \hat{\psi})^a\vee 
Z_{j\alpha}\Big]_0, \quad \overline{\psi}^{(j)a}_\alpha =\frac{1}{4}
\Big[(\overline{\psi} + \hat{\overline{\psi}})^a\vee Z_{\alpha j}\Big]_0.
%\label{definitions of gauge and fermion fields}
\end{eqnarray*}
Here we have explicitly shown the suffices of gauge algebra to stress the
difference of the gauge and flavor suffices. $[\quad ]_0$ denotes the same
notation as (\ref{psi expansion coefficient}).

In order to introduce the realistic leptons and quarks, we need to identify
those states with the eigenstates of the isospin and hypercharge
corresponding to the graded Lie algebra of the supergroup $SU(2|1)$. We
first consider the lepton multiplet of the electron sector by assuming that
the electron neutrino possesses a small mass according to the recent
neutrino experiments~\cite{neutrino}. Correspondingly we consider quartet
states $\nu_L,e_L,e_R,\nu_R$ for the electron sector which are classified by
the quantum number of the hypercharge $y$, the magnitude and the third
component of isospin $t$ and $t_3$. Denoting the eigenstate as 
$|y, t, t_3 \rangle$, we identify 
\begin{equation}
\begin{array}{rclcrcl}
|\nu_L\rangle &=& \displaystyle\bigg|-1, \frac{1}{2}, \frac{1}{2}\bigg\rangle, 
 &\quad& |e_L\rangle 
 &=& \displaystyle\bigg|-1, \frac{1}{2}, -\frac{1}{2}\bigg\rangle, \\
|e_R\rangle &=& |-2,0,0\rangle, &\quad& |\nu_R\rangle &=& |0,0,0\rangle,
\end{array}
\end{equation}
which corresponds to the representation
% 
%%%%%
\renewcommand{\arraystretch}{1.0}
%%%%%
%
%%% new array
\makeatletter
\def\@arrayacol{\edef\@preamble{\@preamble \hskip .7\arraycolsep}}
\def\array{\let\@acol\@arrayacol \let\@classz\@arrayclassz
\let\@classiv\@arrayclassiv \let\\\@arraycr\def\@halignto{}\@tabarray}
\makeatother
\begin{equation}
T_{i}=\left( 
\begin{array}{@{\,}cc|cc}
& 
\mbox{ \hspace{-7.6mm} \raisebox{-2.5mm}[0pt][0pt]
                  { $ \displaystyle \ \frac{\sigma_{i}}{2} $ } } & 0 & 0 \\ 
&  & 0 & 0 \\ \hline
0 & 0 & 0 & 0 \\ 
0 & 0 & 0 & 0
\end{array}
\right), \quad 
Y =\left( 
\begin{array}{@{\,}cc|cc}
-1 & 0 & 0 & 0 \\ 
0 & -1 & 0 & 0 \\ \hline
0 & 0 & -2 & 0 \\ 
0 & 0 & 0 & 0
\end{array}
\right).  
\label{original even generator}
\end{equation}
The above identification of the electron sector satisfies the
Nishijima-Gellmann relation: electric charge $Q=T_3 + \frac{Y}{2}$.

The matrix elements corresponding to the odd counterpart of $SU(2|1)$
generators which can be read from the relations in the appendix are given by 
\begin{eqnarray}
\label{original odd generator}
\Sigma_{+}=\left( 
\begin{array}{@{\,}cc|cc}
0 & 0 & 0 & \gamma \\ 
0 & 0 & 0 & 0 \\ \hline
0 & \beta & 0 & 0 \\ 
0 & 0 & 0 & 0
\end{array}
\right), &\quad&
\Sigma_{-}=\left( 
\begin{array}{@{\,}cc|cc}
0 & 0 & 0 & 0 \\ 
0 & 0 & 0 & \gamma \\ \hline
-\beta & 0 & 0 & 0 \\ 
0 & 0 & 0 & 0
\end{array}
\right),  \nonumber \\
&&  \\        
\Sigma^{\prime}_{+}=\left( 
\begin{array}{@{\,}cc|cc}
0 & 0 & \epsilon & 0 \\ 
0 & 0 & 0 & 0 \\ \hline
0 & 0 & 0 & 0 \\ 
0 & \alpha & 0 & 0
\end{array}
\right), &\quad& 
\Sigma^{\prime}_{-}=\left( 
\begin{array}{@{\,}cc|cc}
0 & 0 & 0 & 0 \\ 
0 & 0 & -\epsilon & 0 \\ \hline
0 & 0 & 0 & 0 \\ 
\alpha & 0 & 0 & 0
\end{array}
\right),  \nonumber   
\end{eqnarray}
%
%%% new array
\makeatletter
\def\@arrayacol{\edef\@preamble{\@preamble \hskip .2\arraycolsep}}
\def\array{\let\@acol\@arrayacol \let\@classz\@arrayclassz
\let\@classiv\@arrayclassiv \let\\\@arraycr\def\@halignto{}\@tabarray}
\makeatother
where four free parameters $\alpha, \beta,\gamma,\epsilon$ are introduced.
To be consistent with the generator relations: 
$$
\{ \Sigma_{\pm}, \Sigma^{\prime}_{\pm} \}=\sqrt{2}T_{\pm}, \quad 
\{ \Sigma_{\pm}, \Sigma^{\prime}_{\mp} \}=\pm T_{3}+\frac{1}{2}Y,
$$
these parameters must satisfy 
\begin{equation}
\alpha \gamma + \beta \epsilon = 1, \quad \alpha \gamma =0, \quad
\beta \epsilon =1.  
\label{parameter relations 1}
\end{equation}
We can then take the following choice: 
\begin{equation}
\epsilon =\frac{1}{\beta}, \quad \alpha=0,  
\label{parameter choice 1}
\end{equation}
for the odd generators in (\ref{original odd generator}).

Identifying $\psi^{(j)a} =(\nu_L,e_L,e_R, \nu_R)^t$ and introducing the above
generators in the matter action $S_M$, we obtain the Higgs-fermion coupling
action of the Weinberg-Salam model. The quartet difference of the electron
sector can be denoted by the suffix $a$ while the flavor suffix $(j)$ is
neglected here. We simply consider that there are 4 copies of electron
sector and thus we neglect the physical significance of the result of the
Dirac-K\"ahler formulation.

In order to obtain the mass term of the fermion sector we need some care
since the odd generators are usually not Hermitian. We can, however, use an
automorphism $c$ which is special to the Lie superalgebra $SU(2|1)$ and
corresponds to the charge conjugation. In other words there is an equivalent
representation of the algebra which induces charge conjugate states~\cite
{Coq4,Rittenberg2,Rittenberg3}, 
%
%%%%%
\renewcommand{\arraystretch}{1.6}
%%%%%
%
\begin{equation}
\begin{array}{c}
(T_{\pm})^c=T_{\mp}, \quad (T_3)^c=-T_3, \quad (Y)^c=-Y,  \\
(\Sigma_{\pm})^c=\pm \Sigma^{\prime}_{\mp}, \quad
(\Sigma^{\prime}_{\pm})^c=\pm\Sigma_{\mp}.
\end{array}  
\label{automorphism}
\end{equation}
We then identify the charge conjugate states of the electron sector as 
\begin{equation}
\begin{array}{rclcrcl}
|e_R^c\rangle &=& \displaystyle\bigg|1, \frac{1}{2}, \frac{1}{2}\bigg\rangle, 
&\quad&  -|\nu_R^c\rangle 
&=& \displaystyle\bigg|1, \frac{1}{2}, -\frac{1}{2}\bigg\rangle,
\nonumber \\
|e_L^c\rangle &=& |2,0,0\rangle, &\quad& |\nu_L^c\rangle &=& |0,0,0\rangle,
\end{array}
\end{equation}
which are consistent with the Nishijima-Gellmann relation: $Q^c=T^c_3
+\frac{Y^c}{2}$ with the following even generators: 
%
%%%%%
\renewcommand{\arraystretch}{1.0}
%%%%%
%
%%% new array
\makeatletter
\def\@arrayacol{\edef\@preamble{\@preamble \hskip .7\arraycolsep}}
\def\array{\let\@acol\@arrayacol \let\@classz\@arrayclassz
\let\@classiv\@arrayclassiv \let\\\@arraycr\def\@halignto{}\@tabarray}
\makeatother
\begin{eqnarray}
T_{i}^c=\left( 
\begin{array}{@{\,}cc|cc}
& 
\mbox{ \hspace{-7.6mm} \raisebox{-2.5mm}[0pt][0pt]
                  { $ \displaystyle \ \frac{\sigma_{i}}{2} $ } } & 0 & 0 \\ 
&  & 0 & 0 \\ \hline
0 & 0 & 0 & 0 \\ 
0 & 0 & 0 & 0
\end{array}
\right), \quad Y^c =\left( 
\begin{array}{@{\,}cc|cc}
1 & 0 & 0 & 0 \\ 
0 & 1 & 0 & 0 \\ \hline
0 & 0 & 2 & 0 \\ 
0 & 0 & 0 & 0
\end{array}
\right).  
\label{charge conjugate even generator}
\end{eqnarray}
We can now identify the charge conjugate electron sector $\psi^{c(j)a}
=(e_R^c,-\nu_R^c,e_L^c,\nu_L^c)^t$, where the flavor suffix $(j)$ is again
neglected here.

This choice of the even generators $T_i^c$ does not directly satisfy the
relations (\ref{automorphism}) with respect to the original generators $T_i$
of (\ref{original even generator}). In this representation of charge
conjugation matrix, the first and second suffices of row and column are
interchanged with respect to the naive charge conjugation matrix.
Correspondingly the odd generators of charge conjugation matrix $\Sigma^c$
can be obtained by a similar change 
\begin{eqnarray}
\Sigma_{+}^c&=&\left( 
\begin{array}{@{\,}cc|cc}
0 & 0 & -\frac{1}{\beta} & 0 \\ 
0 & 0 & 0 & 0 \\ \hline
0 & 0 & 0 & 0 \\ 
0 & 0 & 0 & 0
\end{array}
\right), \quad 
\Sigma_{-}^c=\left( 
\begin{array}{@{\,}cc|cc}
0 & 0 & 0 & 0 \\ 
0 & 0 & -\frac{1}{\beta} & 0 \\ \hline
0 & 0 & 0 & 0 \\ 
0 & 0 & 0 & 0
\end{array}
\right),  \nonumber \\
&& \\
{\Sigma^{\prime}}^c_{+}&=&\left( 
\begin{array}{@{\,}cc|cc}
0 & 0 & 0 & \gamma \\ 
0 & 0 & 0 & 0 \\ \hline
0 & -\beta & 0 & 0 \\ 
0 & 0 & 0 & 0
\end{array}
\right), \quad 
{\Sigma^{\prime}}^c_{-}=\left( 
\begin{array}{@{\,}cc|cc}
0 & 0 & 0 & 0 \\ 
0 & 0 & 0 & -\gamma \\ \hline
-\beta & 0 & 0 & 0 \\ 
0 & 0 & 0 & 0 
\end{array}
\right).  \nonumber 
\label{charge conjugate odd generator}
\end{eqnarray}
%
%%% new array
\makeatletter
\def\@arrayacol{\edef\@preamble{\@preamble \hskip .2\arraycolsep}}
\def\array{\let\@acol\@arrayacol \let\@classz\@arrayclassz
\let\@classiv\@arrayclassiv \let\\\@arraycr\def\@halignto{}\@tabarray}
\makeatother

The mass term of the electron sector can be obtained by adding the original
fermionic mass term and the charge conjugate fermionic mass term with
Hermitian conjugate, 
\begin{eqnarray}
S_{\scriptsize\mbox{mass}} &=& \int \Big[-\frac{i}{4}(\overline{\psi}
+\hat{\overline{\psi}})\vee (m + \hat{A})\vee (\psi+\hat{\psi})  \nonumber \\
&&\hspace{7mm} -\frac{i}{4}(\overline{\psi}+\hat{\overline{\psi}})^c\vee (m^c 
+ \hat{A}^c)\vee (\psi+\hat{\psi})^c + \mbox{h.c.} \Big]*1  \nonumber \\
&=&\int d^4x\Big[\sqrt{6|y|}v (-\frac{1}{\beta}\overline{e}e + \gamma \ 
\overline{\nu}{\nu}) +\cdots \Big],  
\label{lepton mass term}
\end{eqnarray}
where the terms $[\cdots]$ include Higgs-lepton Yukawa coupling terms. We
have chosen $v$ to be real and $y<0$ for the spontaneously broken phase. As
we can see from (\ref{lepton mass term}), the electron mass and neutrino
mass are given by $-\frac{1}{\beta} \sqrt{6|y|}v$ and $\gamma \sqrt{6|y|}v$,
respectively. These masses are then controlled by the free parameters $\beta$
and $\gamma$.

The quark sector can be treated in a similar way as the lepton sector. In
the case of the quark sector we assign the quartet of $u$, $d$ quark 
eigenstates as 
%
%%%%%
\renewcommand{\arraystretch}{2.0}
%%%%%
%
\begin{equation}
\begin{array}{rclcrcl}
|u_L\rangle 
&=& \displaystyle\bigg|\frac{1}{3}, \frac{1}{2}, \frac{1}{2}\bigg\rangle, 
&\quad&  
|d_L\rangle 
&=&\displaystyle \bigg|\frac{1}{3}, \frac{1}{2}, -\frac{1}{2}\bigg\rangle,  \\
|u_R\rangle
&=& \displaystyle\bigg|\frac{4}{3}, 0, 0 \bigg\rangle, 
&\quad& |d_R\rangle
&=& \displaystyle\bigg|-\frac{2}{3}, 0, 0\bigg\rangle, 
\label{quark eigenstate}
\end{array}
\end{equation}
where the Nishijima-Gellmann relation leads
%
%%%%%
\renewcommand{\arraystretch}{1.2}
%%%%%
%
%%% new array
\makeatletter
\def\@arrayacol{\edef\@preamble{\@preamble \hskip .7\arraycolsep}}
\def\array{\let\@acol\@arrayacol \let\@classz\@arrayclassz
\let\@classiv\@arrayclassiv \let\\\@arraycr\def\@halignto{}\@tabarray}
\makeatother
\begin{equation}
Q = T_3 + \frac{Y}{2} 
  = \left( 
\begin{array}{@{\,}cc|cc}
\frac{2}{3} & 0 & 0 & 0 \\ 
0 & -\frac{1}{3} & 0 & 0 \\ \hline
0 & 0 & \frac{2}{3} & 0 \\ 
0 & 0 & 0 & -\frac{1}{3}
\end{array}
\right),  
\label{quark charge assignment}
\end{equation}
for the hypercharge generator 
\begin{equation}
Y = \left( 
\begin{array}{@{\,}cc|cc}
\frac{1}{3} & 0 & 0 & 0 \\ 
0 & \frac{1}{3} & 0 & 0 \\ \hline
0 & 0 & \frac{4}{3} & 0 \\ 
0 & 0 & 0 & -\frac{2}{3}
\end{array}
\right).  
\label{quark hypercharge}
\end{equation}
%
%%% new array
\makeatletter
\def\@arrayacol{\edef\@preamble{\@preamble \hskip .2\arraycolsep}}
\def\array{\let\@acol\@arrayacol \let\@classz\@arrayclassz
\let\@classiv\@arrayclassiv \let\\\@arraycr\def\@halignto{}\@tabarray}
\makeatother
It is important to note here that the lepton hypercharge generator and the
quark hypercharge generator both satisfy Str$Y=0$. This is one of the most
elegant points of the $SU(2|1)$ graded Lie algebra that the natural choice
of the hypercharge generator leads to satisfy the Nishijima-Gellmann
relation. This point was stressed by Ne'eman~\cite{Neeman} and Coquereaux
et al.~\cite{Coq4}.

%%%%%%%%%%%%%%%%%%% 

\section{Summary and discussions}

\setcounter{equation}{0}
\setcounter{footnote}{0}

We have extended the formulation of generalized gauge theory of Chern-Simons
type and topological Yang-Mills type actions to Yang-Mills type actions. In
extending the generalized gauge theory formulation to Yang-Mills type
actions we needed to introduce the cup product of differential forms, which
breaks the topological nature of the generalized gauge theory since the cup
product includes Hodge star operation and thus needs a particular choice of
the metric.

As an application of the generalized Yang-Mills action we have introduced
only 0-form and 1-form gauge fields which correspond to Higgs and gauge
fields, respectively. Since the graded Lie algebra is the natural gauge
algebra of the generalized gauge theory, we have chosen the graded Lie
algebra of $SU(2|1)$ supergroup. We can then derive the Weinberg-Salam
model from the generalized gauge theory formulation where the quaternion
plays the role of the two by two matrix representing the discrete two points
of noncommutative geometry formulation $\grave{\mathit{a}}$ \textit{la}
Connes. Matter fermions are formulated by Dirac-K\"ahler fermion
formulation by employing fermionic differential forms. In this way the
Weinberg-Salam model has been formulated purely by differential forms.

In the formulation of the generalized Yang-Mills action, only the 0-form
gauge parameter appears as an effective gauge parameter. The generalized
gauge invariance is lost for the higher form gauge parameters due to the
break down of the associativity by the mixture of the product between the
wedge product and the cup product. If we, however, formulate the generalized
Yang-Mills action introducing only the cup product from the beginning, we
can construct an interesting Yang-Mills action which lacks differential
operator but possesses the gauge invariance of all the degrees of
differential forms of the generalized gauge parameters. If we define the
generalized curvature by anticommutator with the cup product instead of the
standard wedge product as 
$$
\mathcal{F} \equiv \frac{1}{2} \{ \mathcal{D} + \mathcal{A}, \mathcal{D} + 
\mathcal{A} \}_\vee  
%\label{gc with cup product}
$$
there appear unwanted derivative terms. We then define a new curvature
neglecting the derivative terms in this definition. We can now define the
generalized Yang-Mills action with matter fermions without derivative terms 
$$
S = \int \Big[\hbox{Tr}[ {\mathcal{A}}\vee {\mathcal{A}} 
\vee{\mathcal{A}}\vee{\mathcal{A}}] + \overline{\Psi} \vee 
{\mathcal A}\vee\Psi\Big]_{\hbox{{\bf 1}}}*1,  
%\label{generalized reduced model}
$$
which is invariant under the following generalized gauge transformation
since the associativity is recovered now, 
\begin{eqnarray*}
&&\delta {\mathcal{A}} = [{\mathcal{A}}, {\mathcal{V}}]_\vee, \\
&&\delta \Psi = - {\mathcal{V}}\vee \Psi, \quad 
\delta \overline{\Psi} = \overline{\Psi}\vee {\mathcal{V}}.   
\end{eqnarray*}
The gauge parameters include all the degrees of differential forms as in 
(\ref{ggp2}) and should satisfy the consistency constraints 
(\ref{consistency condition}). We need to choose one of the quaternion algebra 
(\ref{quaternion algebra 1}) or (\ref{quaternion algebra 2}) 
in this formulation.
As we can see that this action has similar structure 
as the reduced model~\cite{Kawai} yet this is not the reduced model 
in the standard sense since the gauge fields and parameters have 
spacetime dependence. In the present formulation fermions are in 
the fundamental representation, we can, however, introduce adjoint 
representations as well.
We expect that this model may have essential connection with the formulation 
of the generalized gauge theory formulation on the simplicial lattice.

In the formulation of matter fermion of the Weinberg-Salam model by the
Dirac-K\"ahler fermion formulation, the flavor or, it is better to say,
family suffix naturally appears but this family suffix has been neglected in
the formulation. In our real nature the number of family is three instead of
four which is the natural consequence of the Dirac-K\"ahler fermion
formulation. We expect that the family suffix and the suffix of the gauge
algebra would be interrelated in the real unified theory of the formulation
which we are hoping to derive.

In fact there is a toy model of this type that the gauge algebra and the
Dirac-K\"ahler matter fermion and supersymmetry are naturally
interrelated. This is the formulation of two-dimensional version of the
generalized topological Yang-Mills action with the instanton 
gauge fixing~\cite{KT}. In this formulation $N=2$ super Yang-Mills 
action comes out naturally and the matter fermions appears from ghosts 
of quantization via twisting mechanism just like in four-dimensional 
topological field theory~\cite{Witten}. The twisting mechanism and 
the Dirac-K\"ahler fermion formulation are essentially related through 
supersymmetry.

We hope that the generalized gauge theory presented in this paper may play
an essential role in unifying the standard model and quantum gravity on the
lattice.

\vskip 1cm 
%%%%%%%%%%
\noindent{\Large \textbf{Acknowledgments}}

One of the authors (N. K.) would like to thank D.S. Hwang and C.-Y. Lee for
useful discussions at the very early stage of this work. This work is
supported in part by the Grant-in-Aid for Scientific Research from the
Ministry of Education, Science and Culture of Japan under the grant number
09640330.

\vskip 1cm

\begin{center}
{\Large \textbf{Appendix}\\[0pt]}
\end{center}

The matrix elements of $SU(2|1)$ generators~\cite
{Rittenberg1,Rittenberg2,Rittenberg3,Coq4} are given by 
\begin{eqnarray*}
Y|y, t, t_3\rangle 
&=& y|y, t, t_3\rangle, \\
T_3|y, t, t_3\rangle
&=& t_3|y, t, t_3\rangle,  \\
T_\pm |y, t, t_3\rangle 
&=& \sqrt{\frac{(t\mp t_3)(t\pm t_3+1)}{2}}|y, t, t_3\pm1\rangle,  \\
\Sigma_\pm |y,t,t_3\rangle
&=& \pm\beta\sqrt{t\mp t_3} \bigg|y-1, t-\frac{1}{2}, t_3\pm\frac{1}{2}
                            \bigg\rangle,  \\
\Sigma^{\prime}_\pm |y,t,t_3\rangle
&=& \alpha\sqrt{t\mp t_3} \bigg|y+1, t-\frac{1}{2}, t_3\pm\frac{1}{2}
                          \bigg\rangle,  \\
\Sigma_\pm \bigg|y-1, t-\frac{1}{2}, t_3\bigg\rangle 
&=& 0,  \\
\Sigma^{\prime}_\pm \bigg|y-1, t-\frac{1}{2}, t_3\bigg\rangle 
&=& \pm\epsilon\sqrt{t\pm t_3+\frac{1}{2}}
    \bigg|y, t, t_3\pm \frac{1}{2}\bigg\rangle  \\
&&+\zeta\sqrt{t\mp t_3-\frac{1}{2}}
  \bigg|y, t-1, t_3\pm\frac{1}{2}\bigg\rangle, \\
\Sigma_\pm \bigg|y+1, t-\frac{1}{2}, t_3\bigg\rangle 
&=& \gamma\sqrt{t\pm t_3+\frac{1}{2}}\bigg|y, t, t_3\pm \frac{1}{2}
                                     \bigg\rangle \\
&&\pm\delta\sqrt{t\mp t_3-\frac{1}{2}}\bigg|y, t-1, t_3\pm\frac{1}{2}
                                      \bigg\rangle,  \\
\Sigma^{\prime}_\pm\bigg|y+1, t-\frac{1}{2}, t_3\bigg\rangle 
&=& 0,  \\
\Sigma_\pm|y, t-1, t_3\rangle 
&=& \omega\sqrt{t\mp t_3} \bigg|y-1, t-\frac{1}{2}, t_3\pm\frac{1}{2}
                          \bigg\rangle,  \\
\Sigma^{\prime}_\pm|y,t-1,t_3\rangle 
&=& \pm\tau\sqrt{t\mp t_3} \bigg|y+1, t-\frac{1}{2}, t_3\pm\frac{1}{2}
                           \bigg\rangle.  
\label{matrix element of su(2|1) algebra} 
\end{eqnarray*}
%

%%%%%%%%%%%%%%%%%%%%

\end{document}